\documentclass[twocolumn,journal,final]{IEEEtran}
\usepackage{algorithmicx,fullpage,epstopdf}
\usepackage{algorithm}
\usepackage{amsmath,epsfig,subfigure,paralist,verbatim,tikz}
\usetikzlibrary{arrows}
\usepackage{color}
\usepackage{amsfonts}
\usepackage{amssymb}
\usepackage{epsfig,psfrag}

\newtheorem{theorem}{Theorem}[section]

\newcommand{\Y}{\mathbb{Y}} \newcommand{\X}{\mathbb{X}}
 
\def\S{{\mathbf S}}

\newcommand{\tp}{P}

\newcommand{\belief}{{\pi}}
\newcommand{\tbelief}{\pi^0}

\newcommand{\priv}{\eta}

\newcommand{\A}{\mathcal{A}}

\newcommand{\beq}{\begin{equation}}
\newcommand{\eeq}{\end{equation}}
\newcommand{\E}{\mathbf{E}}

\newcommand{\history}{\mathcal{H}}
\newcommand{\full}{\mathcal{F}}

\newcommand{\argmax}{\operatorname{argmax}}
\newcommand{\argmin}{\operatorname{argmin}}

\newcommand{\Bs}{R^\pi} 
\newcommand{\ta}{\tilde{a}}
\newcommand{\sigs}{\sigma}
  \def\1{{\mathbf 1}}

\newcommand{\nn}{\nonumber}
\newcommand{\qed} {{$\hfill\blacksquare$}}

\newcommand{\pizero}{\pi_0}

\newcommand{\Ep}{\E^\mu_{\pi_0}}

\newcommand{\Pp}{\mathbf{P}^\mu_{\pi_0}}

\newcommand{\lbelief}{l}

\newcommand{\sigp}{\sigma}
\newcommand{\discount}{\rho}

    \newcommand{\Ts}{T}
    \newcommand{\ca}{c_a}
\newcommand{\p}{\prime}
    \def\S{{S}} 

\begin{document}
\title{Interactive Sensing in Social Networks}

\author{Vikram~Krishnamurthy,~\IEEEmembership{Fellow,~IEEE},  H. Vincent Poor,~\IEEEmembership{Fellow,~IEEE} %
\thanks{V. Krishnamurthy (e-mail: vikramk@ece.ubc.ca) is with the Department of Electrical and Computer
Engineering, University of British Columbia, Vancouver, V6T 1Z4, Canada. H.V. Poor (email poor@princeton.edu) is with the Department of Electrical
Engineering, Princeton University, Princeton, NJ,  08544, USA.  

This work was supported in part by the Canada Research Chairs program, NSERC Canada, in part by the U.S.\ Army Research
Office under MURI Grant W911NF-11-1-0036  and  in part by the U.\S.\ National Science Foundation under
Grant CNS-09-05086.
}

}

\maketitle

\begin{abstract}

This paper presents models and algorithms for interactive sensing in social networks where individuals act as sensors and  the information  exchange between individuals
is exploited to optimize sensing. Social learning is used 
  to model the interaction
between individuals that aim to estimate an underlying state of nature.
In this context the following questions are addressed:
 How can self-interested agents  that interact via social learning achieve a tradeoff between
 individual privacy and reputation of the social group?
 How can  protocols be designed to prevent data incest in  online reputation blogs where individuals make recommendations?
How can sensing by  individuals that interact with each other be used by a global decision  maker 
to detect changes in the underlying state of nature? 
 When individual agents possess limited sensing, computation and communication capabilities, can a network of agents achieve sophisticated global behavior?  Social and game theoretic learning  are natural settings for addressing these questions. 
 This article presents an overview,  insights and discussion of social learning models
in  the context of 
  data incest propagation, change detection and coordination of decision making. \end{abstract}

\section{Introduction and Motivation}
\label{sec:introduction}

The proliferation of  social media such as  real time microblogging services (Twitter\footnote{On US Presidential election day in 2012, there were 15 thousand tweets per second resulting in 500 million tweets in the day. Twitter can be considered as a real time sensor.}),  online reputation and rating
systems (YELP) together with  app-enabled smartphones,
 facilitate   real time sensing of social activities, social patterns  and  behavior. 

{\em Social sensing}, also called participatory sensing \cite{EP06,AA11,CEL08,BEH06}, is defined as a process where physical sensors present in mobile devices such as GPS are used to infer social relationships and human activities.
 In this paper,  we work at a higher level of abstraction.
We use the term {\em social sensor} or
{\em  human-based sensor} to denote an agent that provides information about its environment  (state of nature) on a social network  after
interaction  with other agents. Examples of  such social sensors include Twitter posts, Facebook status updates, and ratings on online reputation systems like YELP and Tripadvisor.
Such social sensors go beyond physical sensors for social sensing. For example  \cite{RMZ11}, user opinions/ratings (such as the quality of a restaurant) are available on Tripadvisor
but  are difficult to measure via  physical sensors.
 Similarly, future situations revealed by the Facebook status of a user  are impossible to predict using  physical sensors.

Statistical inference using social sensors is relevant in a variety of applications
including  localizing special events  for targeted advertising \cite{LS10,CCL10}, marketing \cite{TBB10},
localization of natural disasters \cite{SOM10}  and  predicting sentiment  of investors in financial markets  \cite{BMZ11,PL08}.
It is demonstrated in \cite{AH10}  that  models built from the rate of tweets  for particular products can outperform  market-based predictors.
%
However, social sensors  present unique challenges from a statistical estimation point of view.
First,  social sensors  interact with and influence other social sensors. For example, ratings posted on online reputation systems strongly influence the behaviour of  individuals.\footnote{It is reported
 in \cite{IMS11} that  81\% of hotel managers  regularly check Tripadvisor reviews.  \cite{Luc11} reports that a one-star increase in the Yelp rating maps to 5-9 \% revenue increase. \label{foot}}
Such interacting sensing can result in  non-standard information patterns due to correlations introduced by the structure of the underlying social network.
Second, due to privacy
reasons and time-constraints,
social sensors typically do not reveal  raw observations of the underlying state of nature.
Instead, they reveal their decisions 
(ratings, recommendations, votes) which can be viewed as a low resolution (quantized)   function of their raw measurements and interactions with other social sensors.

As is apparent from the above discussion,
there is strong motivation to construct mathematical models that capture the dynamics of interactive sensing involving social sensors. Such models  facilitate understanding the dynamics of information
flow in social networks and therefore  the design of algorithms that can exploit these dynamics to estimate the underlying state of nature.

In this paper,
{\em social learning} \cite{Ban92,BHW92,Cha04}  serves as a useful  mathematical abstraction for modelling  the interaction of social  sensors.
Social learning in  multi-agent systems  seeks to answer the following question:
\begin{quote}  {\em How do decisions made by agents affect decisions made by subsequent agents?}  \end{quote}
  In  social learning,  each agent chooses its action by  optimizing
its local utility function. 
  Subsequent agents then use their private observations together with the actions of previous agents
 to estimate (learn) an underlying state.
The setup is fundamentally different to classical signal processing in which sensors use  noisy observations to compute estimates - in social learning agents
use  noisy observations together with decisions made by previous agents, to estimate the underlying state of nature.

In the last decade, social learning has been used widely in economics, marketing, political science and sociology  to model the behavior of financial markets, 
crowds, social groups and social networks; see \cite{Ban92,BHW92,AO11,Cha04,LADO07,ADLO08} and numerous references therein. Related models have been studied in the context of sequential decision making in
information theory \cite{CH70,HC70}  and statistical signal processing \cite{CSL13,KP13} in the electrical engineering literature.

 Social learning models for interactive  sensing can predict  unusual behavior. Indeed, a key result in social learning of an underlying random variable is that rational agents eventually herd \cite{BHW92}, that is, they eventually end up choosing the same action irrespective of their private observations. As a result, the actions
contain no information about the private observations and so the Bayesian estimate of the underlying random variable freezes. For a multi-agent  sensing system,  such behavior
 can be undesirable, particularly if individuals
 herd and make incorrect decisions. 

\subsection*{Main Results and Organization}

 In the context of social learning models for interactive  sensing, the  main ideas  and organization of this paper are as follows:

\noindent
1. {\em Social Learning Protocol}:
Sec.\ref{sec:classicalsocial} presents a formulation and survey of the classical Bayesian social learning model which forms the mathematical basis for modelling interactive  sensing amongst humans.
 We illustrate  the social learning model  in the context of Bayesian signal processing (for easy access to an electrical engineering
audience). 
We then address how self-interested agents performing social learning can achieve  useful behavior in terms of
 optimizing a social welfare function. Such problems are motivated by privacy
 issues in sensing. If an agent reveals less information
  in its decisions,
 it maintains its privacy; on the other hand as part of a social group it has an incentive to optimize a social welfare
 function that helps estimate the state of nature.
\\
2. {\em Data Incest in Online Reputation Systems}:  
Sec.\ref{sec:incest} deals with the question:
How can data incest (misinformation propagation) be prevented in online reputation blogs where social sensors  make recommendations? 

In the classical social learning model,  each agent  acts once in a pre-determined order.
However, in online reputation systems  such as Yelp or Tripadvisor which maintain logs of votes (actions)  by agents, social learning takes place with information exchange over a loopy graph  (where the agents form the vertices of the graph).
Due to the loops in the information exchange graph,
{\em data incest} (mis-information) can propagate: 
Suppose an agent wrote  a poor rating of a restaurant on a social media site.  Another agent is influenced by this rating, visits the restaurant, and then also gives  a poor rating on the social media site.
  The first agent visits the social media site  and notices that another agent has also given the restaurant a poor rating - this double confirms her rating and she enters another poor rating. 
  
   In a fair 
  reputation
  system,  such ``double counting" or data incest should have been prevented by making the first agent  aware that the rating of the second agent was influenced by her own rating.
Data incest results in a  bias in the estimate of state of nature. 
 How can automated protocols be designed to prevent data incest and thereby maintain a fair\footnote{Maintaining fair reputation systems has  financial implications as is apparent from
 footnote \ref{foot}.
 }  online reputation system? Sec.\ref{sec:incest} describes how the administrator of a social network can maintain an unbiased (fair) reputation system.
\\
3.   {\em  Interaction of Local and  Global Decision Makers for Change Detection}:  Sec.\ref{sec:socialc} deals with the question:  In sensing where individual agents perform social learning to estimate an underlying state of nature, how can changes in the  state of nature be detected?
Sec.\ref{sec:socialc}  considers a  sensing problem
 that involves  change detection. 
 Such   sensing problems arise in a variety of applications such as financial trading where individuals react to financial shocks
\cite{AS08};  marketing and advertising \cite{Pin06,Pin08} where consumers
react to a new product;  and  localization of
natural disasters (earthquake and typhoons) \cite{SOM10}.
 
For example, consider measurement of the adoption of a  new  product using a micro-blogging platform like twitter. The adoption of the technology diffuses through the market but its effects can only be observed through the tweets of select members of the population. These selected members act as  sensors for the parameter of interest.
 Suppose the state of nature  suddenly  changes due to  a sudden market shock or presence of a new competitor.
Based on the local actions of the multi-agent system that is performing social learning, a global decision maker (such as a market monitor or technology manufacturer) needs to decide  whether or not to declare if a change has occurred.
How can the global decision maker achieve such change detection to minimize a cost function comprised of false alarm rate and delay penalty? The local and global decision makers
interact, since
the local decisions determine the posterior distribution of subsequent agents which determines the global decision (stop or continue) which determines subsequent  local decisions.
 We show that this social learning based change detection problem leads to unusual behavior. 
The optimal decision policy of the stopping time problem has multiple thresholds.    This  is unusual: 
 if it is optimal to declare that a change has occurred based on the posterior probability of change, it may not be optimal to declare a change when the posterior probability of change is higher!
\\
4. {\em Coordination of Decisions as a Non-cooperative Game}: No discussion on social learning would be complete
without mentioning game-theoretic methods.
A large body of research on social networks has been devoted to the diffusion of information (e.g., ideas, behaviors, trends) \cite{Gra78,GLM01}, and particularly on finding a set of target nodes so as to maximize the spread of a given product \cite{MR07,Che09}. Often  customers end up choosing a specific  product among several competitors.
A  natural approach to model this competitive process is via the use of non-cooperative game theory \cite{BKS07,AM11}.

Game theory has traditionally been used in economics and social sciences with a focus on fully rational interactions where strong assumptions are made on the
information patterns available to individual agents. In comparison,  social sensors are  agents with partial information and it is the dynamic
interactions between agents that is of interest. This motivates the need for game theoretic learning models for agents interacting in social networks.

Sec.\ref{sec:game} deals with the question:
 When individuals  are self-interested and possess limited sensing, computation and communication capabilities,
 can  a network
 (social group)
of  sensors whose utility functions interact achieve sophisticated global behavior?
In Sec.\ref{sec:game},
 we  discuss a  non-cooperative game theoretic learning approach for adaptive decision making in social networks. This can be viewed as a non-Bayesian version
 of social learning, The aim is to ensure that all agents eventually
 choose actions from a common polytope of randomized strategies - namely, the set of correlated equilibria of a non-cooperative game.
Correlated equilibria are a generalization of Nash equilibria and were introduced by Aumann~\cite{Aum87}.\footnote{Aumann's  2005 Nobel prize
in economics press release  reads:
``Aumann also introduced a new equilibrium concept, correlated equilibrium, which is
weaker than Nash equilibrium, the solution concept developed by John Nash, an economics
laureate in 1994. Correlated equilibrium can explain why it may be advantageous for negotiating
parties to allow an impartial mediator to speak to the parties either jointly or separately ..."}

\subsection*{Perspective}

The social learning and game-theoretic learning formalisms mentioned above
 can be used either as descriptive tools, to predict the outcome of complex interactions amongst agents in   sensing, or as  prescriptive tools, to design social networks and sensing
 systems around given interaction rules.
Information aggregation, misinformation propagation
and privacy are important issues in  sensing using social sensors. 
In this paper, we  treat these issues in a highly stylized manner so as to provide easy accessibility to an electrical engineering audience.
The underlying  tools
used in this paper are widely used by  the electrical engineering research community in the areas of signal processing, control, information theory and network communications.

In Bayesian estimation,  
the twin effects of social learning (information aggregation with interaction amongst agents) and data incest 
(misinformation propagation) lead to non-standard information patterns in estimating the underlying state of nature.
 Herding occurs when the public belief overrides the private observations and thus actions of agents are independent of their private observations.
Data incest results in bias in the public belief as a consequence of the unintentional re-use of identical actions in the formation of public belief in social learning; the information gathered by each agent is mistakenly considered to be independent. This results in overconfidence and bias in estimates of the state of nature.

Privacy issues impose important constraints on  social sensors. Typically, individuals are not willing to disclose private observations. Optimizing interactive  sensing with privacy constraints is an important problem. Privacy and trust pose conflicting requirements on human-based sensing:
 privacy requirements  result in  noisier measurements or lower resolution actions, while maintaining a high degree
 of trust (reputation) requires accurate measurements.  Utility functions, noisy private measurements and quantized actions are essential
 ingredients of the  social and game-theoretic  learning models presented in this paper 
that facilitate
modelling this tradeoff between reputation and privacy.

The literature in the areas of social learning, sensing and networking is extensive. Due to page restrictions, in each of the following sections, we provide only a brief review of relevant works.
Seminal books in social networks include \cite{Veg07,Jac10}. The book \cite{Cha04} contains a complete treatment of social learning models with several remarkable insights.
For further references, we refer the reader to \cite{KMY08,MKZ08,Kri08,Kri11,Kri12}.
In \cite{Har05}, a  nice description is given of  how, if individual agents deploy simple heuristics, the global
system behavior can achieve 
"rational" behavior. The related problem of achieving 
{\it coherence} (i.e., agents eventually choosing the same action or the same decision policy) among disparate sensors of decision agents without
cooperation has also witnessed intense research; see \cite{PKP09} and \cite{WKP11}.
 Non-Bayesian social learning models are also studied in \cite{EF93,EF95}.

\section{Multi-agent  Social Learning} 
   \label{sec:classicalsocial}

This section starts with a brief description of the classical social learning model. 
In this paper, we use social learning as the mathematical basis for modelling interaction of social sensors.
A key result in social learning is that rational agents eventually herd, that is, they pick the same
 action irrespective of their private observation and social learning stops.  To delay the effect of herding,  and thereby enhance social learning,  Chamley \cite{Cha04}  (see also \cite{SS97} for related work) 
 has proposed a novel  constrained optimal social learning protocol.
 We review this protocol which is formulated as a  sequential stopping time problem.
 We show
  that the  constrained optimal social learning proposed by Chamley \cite{Cha04}
  has a  threshold switching curve  in the space of public belief states.
Thus the global decision to stop   can be implemented efficiently in a social learning model.

\subsection{Motivation: What is social learning?}\label{sec:herd}

We start with
a brief description of the `vanilla'\footnote{In typical formulations of social learning, the underlying state is assumed to be a random variable and not a Markov chain. Our description below is given in terms of a Markov chain since we wish to highlight the unusual structure of the social learning filter below to a signal processing reader who is familiar with basic ideas in Bayesian filtering.
Also we are interested in change detection problems in which the change time distribution can be modelled as the absorption time of a Markov chain.} social learning model.
In social learning \cite{Cha04},
 agents estimate the underlying state of nature not only from their local measurements, but also from the actions of previous
agents. (These previous actions  were taken by agents in response to their local measurements; therefore these actions
convey information about the underlying state).
As we will describe below,
the  state estimation update  in social learning has a drastically different
structure compared to the standard optimal filtering recursion and can result in unusual behavior.

Consider a countable  number of agents performing social learning  to estimate the state of an underlying finite state Markov chain $x$.
 Let $\X = \{1,2,\ldots,X\}$ denote a finite state space, $\tp$ the transition matrix and $\pi_0$ the initial distribution of the Markov chain.

Each agent acts once  in a predetermined sequential order indexed by $k=1,2,\ldots$   The index $k$ can also be viewed
as the discrete time instant when agent $k$ acts.  
A multi-agent system seeks to  estimate $x_0$.  Assume at the beginning of iteration $k$,
all agents have access to the public belief $\pi_{k-1}$ defined in  Step (iv) below.
The social learning protocol proceeds as follows
 \cite{BHW92,Cha04}:\\
 (i) {\em Private Observation}: At time $k$,
agent $k$  records a private observation $y_k\in \Y $ 
from the observation distribution $B_{iy} = P(y|x=i)$, $i \in \X$.
Throughout this section we assume that $\Y = \{1,2,\ldots,Y\}$ is finite.
\\
(ii) 
{\em Private Belief}:  Using the public belief $\pi_{k-1} $ available at time $k-1$ (defined in Step (iv) below), agent $k$   updates its private
posterior belief  $\priv_k(i)  =  P(x_k = i| a_1,\ldots,a_{k-1},y_k)$ as the following Bayesian update (this is  the classical Hidden Markov Model
filter \cite{EM02}):
\begin{align}  \label{eq:hmm} \priv_k &= 
\frac{B_{y_k} \tp^\p \pi}{ \mathbf{1}_X^\p B_y \tp^\p \pi}, \quad  
B_{y_k} = \text{diag}(P(y_k|x=i),i\in \X) . 
 \end{align}
 Here $\mathbf{1}_X$ denotes the $X$-dimensional vector of ones, $\eta_k$ is an $X$-dimensional probability mass function (pmf) and $\tp^\p$ denotes transpose of the matrix
 $\tp$.\\
 (iii)   {\em Myopic Action}: Agent  $k$  takes  action $a_k\in \A = \{1,2,\ldots, A\}$ to  minimize its expected cost 
 \begin{multline}  
  a_k =  \arg\min_{a \in \A} \E\{c(x,a)|a_1,\ldots,a_{k-1},y_k\} \\  =\arg\min_{a\in \A} \{c_a^\p\priv_k\}.    \label{eq:myopic}
  \end{multline}
  Here $\ca = (c(i,a), i \in \X)$ denotes an $X$ dimensional cost vector, and $c(i,a)$ denotes the cost  incurred when the underlying state is $i$ and the  agent chooses action $a$.\\
  Agent $k$ then broadcasts its  action $a_k$ to subsequent agents.\\
(iv) {\em Social Learning Filter}:   
Given the action $a_k$ of agent $k$,  and the public belief $\pi_{k-1}$, each  subsequent agent $k' > k$ 
performs social learning to
 compute the public belief $\pi_k$ according to the following ``social learning  filter":\ \beq \pi_k = \Ts(\pi_{k-1},a_k), \text{ where } \Ts(\pi,a) = 
 \frac{\Bs_a \tp ^\p\pi}{\sigs(\pi,a)}, \label{eq:piupdate}\eeq
 and 
$\sigs(\pi,a) = \mathbf{1}_X^\p \Bs_a \tp^\p \pi$ is the normalization factor of the Bayesian update.
In (\ref{eq:piupdate}),  the public belief $\pi_k(i)  = P(x_k = i|a_1,\ldots a_k)$ and $\Bs_a  = \text{diag}(P(a|x=i,\pi),i\in \X ) $ has elements
\begin{align} \label{eq:aprob}
 & P(a_k = a|x_k=i,\pi_{k-1}=\pi) = \sum_{y\in \Y} P(a|y,\pi)P(y|x_k=i) \\
 &   P(a_k=a|y,\pi) = \begin{cases}  1 \text{ if }  c_a^\p B_y \tp^\p \pi \leq c_{\ta}^\p B_y \tp^\p\pi, \; \ta \in \A \\
 0  \text{ otherwise. }  \end{cases}
 \nn \end{align}
The derivation of the social learning filter (\ref{eq:piupdate}) is given in the discussion below.

\begin{figure}
\centering
\scalebox{0.6}{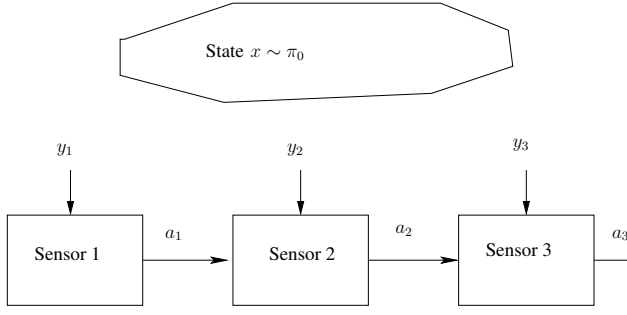}
\caption{Information Exchange Structure in Social Learning}
\label{fig:sl}
\end{figure}

\subsection{Discussion}  Let us pause to give some intuition about  the above social learning protocol. 
\\
1. {\em Information Exchange Structure}: Fig.\ref{fig:sl} illustrates the above social learning protocol in which the information exchange is sequential.  Agents send their hard decisions
(actions) to subsequent agents.
In the  social learning protocol we have assumed that each agent acts once. Another way of viewing the  social learning  protocol
is that there are finitely  many agents that act repeatedly in some pre-defined order. If each
 agent chooses its local decision using the current public belief, then the setting is identical to the social learning
 setup. We also refer the reader to \cite{AO11} for several recent results in social learning
over several types of network adjacency matrices.\\
2.  {\em Filtering with Hard Decisions}: Social learning can be viewed as agents making {\em hard} decision estimates at each time and sending these estimates to subsequent agents.
In conventional Bayesian state estimation, a {\em soft} decision is made, namely, the posterior distribution (or equivalently, observation) is sent
to subsequent agents. For example, if $\A = \X$, and 
the costs are chosen as $c_a = - e_a$ where $e_a$ denotes the unit indicator with $1$ in the $a$-th position, then 
$\argmin_a c_a^\p \pi  = \argmax_a \pi(a)$, i.e.,  the maximum aposteriori probability (MAP) state estimate. For this example, social learning is equivalent
to agents sending the hard MAP estimates to subsequent agents.

 Note that rather than sending a hard decision estimate, if 
each agent chooses its action $a_k = y_k$ (that is agents send their private observations), then
the right-hand side of (\ref{eq:aprob}) becomes $\sum_{y\in \Y} I(y = y_k)  P(y|x_k=i) = P(y_k|x_k=i)$ and so 
 the problem becomes a standard Bayesian filtering problem.
\\
4. {\em Dependence of Observation Likelihood on Prior}: The most unusual feature of the above protocol (to a signal processing audience) is
 the social learning filter (\ref{eq:piupdate}).  In  standard state estimation via a Bayesian filter, the observation likelihood given the state is completely parametrized
 by the observation noise distribution and is functionally independent of the current prior distribution.
 In the social learning filter,
the likelihood of the action given the state (which is denoted by $\Bs_a$) is
an explicit function of the prior $\pi$! 
Not only does the action likelihood depend on the prior, but it is also a discontinuous
function, due to the presence of the $\argmin$ in (\ref{eq:myopic}).  
 \\
5. {\em Derivation of Social Learning Filter}:
The derivation of the social learning filter (\ref{eq:piupdate})  is as follows:  Define the posterior as
 $\belief_k(j) = P(x_k = j|a_1,\ldots, a_k)$. Then
 \begin{align*}
 \belief_k(j) &=  \frac{1}{\sigma(\pi_{k-1},a_k)} P(a_k|x_k=j,a_1,\ldots,a_{k-1}) \\ &  \sum_i P(x_k=j|x_{k-1}=i)
 P(x_{k-1}=i|a_1,\ldots,a_{k-1})  \\
&= \frac{1}{\sigma(\pi_{k-1},a_k)}  \sum_y P(a_k| y_k=y, a_1,\ldots,a_{k-1}) \\ & P(y_k=y|x_k=j)  \sum_i P(x_k=j|x_{k-1}=i) \belief_{k-1}(i)\\
&= \frac{1}{\sigma(\pi_{k-1},a_k)} \sum_y P(a_k|y_k=y,\pi_{k-1}) \\ & \hspace{1cm}P(y_k=y|x_k = j) \sum_i \tp_{ij} \belief_{k-1}(i)
\end{align*}
where the normalization term is 
\begin{multline*} \sigma(\pi_{k-1},a_k)= \sum_j\sum_y P(a_k|y_k=y,\pi_{k-1}) \\ P(y_k=y|x_k = j) \sum_i \tp_{ij} \belief_{k-1}(i). \end{multline*}

  The above social learning protocol and social learning filter (\ref{eq:piupdate}) result in interesting dynamics in state estimation and decision making.
We will illustrate two interesting consequences that are  unusual to an electrical engineering audience:
\begin{compactitem}
\item Rational Agents form herds and information cascades and blindly follow previous agents.
This is
 discussed in Sec.\ref{sec:cascade} below.

\item Making global decisions on change detection in a multi-agent system performing social learning results in multi-threshold behavior.
This is discussed in Sec.\ref{sec:socialc} below.
\end{compactitem}



\subsection{Rational Agents form Information Cascades} \label{sec:cascade}
The first consequence of the unusual nature of  the social learning filter (\ref{eq:piupdate}) is that social learning can result in  multiple
rational 
agents 
taking the same action independently of their observations. To illustrate this behavior,
throughout this subsection, we  assume that  $x$ is a finite state random variable (instead of a Markov chain) with prior distribution $\pi_0$.

We start with the following definitions; see also \cite{Cha04}:
\begin{itemize}
\item  An individual agent $k$ {\em herds} on the public belief $\pi_{k-1}$  if it chooses its action  $a_k = a(\pi_{k-1},y_k)$ in  (\ref{eq:myopic}) independently of its observation
$y_k$.
 \item  A {\em herd of agents} takes place at time $\bar{k}$, if the actions of all agents after time $\bar{k}$ are identical, i.e., $a_k = a_{\bar{k}}$ for all
time  $k > \bar{k}$.
\item An {\em information cascade} occurs at time $\bar{k}$, if the public beliefs of all agents after time $\bar{k}$ are identical, i.e. 
$\pi_k = \pi_{\bar{k}}$ for all $k < \bar{k}$. 
\end{itemize}

Note that if an information cascade occurs, then since the public belief freezes, social learning ceases.
Also from the above definitions it is clear that an information cascade implies a herd of agents, but the reverse is not true; see Sec.\ref{sec:numerical} for an example.

 The following
result which is well known in the economics literature \cite{BHW92,Cha04} states that if agents follow the above social learning protocol, then   after some finite time $\bar{k}$, an
 {\em information cascade} occurs.\footnote{A nice analogy is provided in \cite{Cha04}.
 If I see someone walking down the street with an umbrella, I assume (based on rationality) that he has checked the weather forecast and is carrying
 an umbrella since it might rain. Therefore, I also take an umbrella. So now there are two people walking down the street
 carrying umbrellas. A third person
 sees two people with umbrellas and based on the same inference logic, also takes an umbrella.  Even though each  individual is  rational, 
 such herding behavior  might be irrational  since
 the first person who took the umbrella, may not have checked the weather forecast.
 
 Another example is that of patrons who decide to choose a restaurant. Despite their menu preferences, each patron chooses the restaurant with the
 most  customers. So eventually all patrons herd to one restaurant.
 
 \cite{TBB10}  quotes the following anecdote on user  influence in a social network which can be interpreted as herding: ``... when a popular blogger left his blogging site for a two-week vacation, the site's visitor tally fell, and content produced by three invited substitute bloggers could not stem the decline."}
  The proof follows via an elementary application of the martingale convergence
 theorem.

\begin{theorem}[\cite{BHW92}] 
\label{thm:herd} The social learning protocol described in Sec.\ref{sec:herd}  leads to an information cascade  in finite time
with probability~1. That is there
exists a finite time $\bar{k}$ after which social learning ceases, i.e., public belief $\pi_{k+1} = \pi_k$, $k \geq \bar{k}$, and all agents choose the same action, i.e., $a_{k+1} = a_k$, $k\geq \bar{k}$.
  \qed\end{theorem}

Instead of reproducing the proof, let us give some insight as to why Theorem \ref{thm:herd} holds. It can be shown using martingale methods
that at some finite time $k=k^*$, the agent's probability $P(a_k|y_k,\pi_{k-1})$ becomes  independent of the private observation $y_k$. Then clearly
from (\ref{eq:aprob}), $P(a_k=a|x_k=i,\pi_{k-1}) =   P(a_k=a|\pi)$. Substituting this into the social learning filter (\ref{eq:piupdate}), we see
that $\pi_{k} = \pi_{k-1}$.  Thus after some finite time $k^*$, the social learning filter hits a fixed point and social learning stops. 
As a result, all subsequent agents $k> k^*$ completely disregard their private observations and take the same action $a_{k^*}$,
thereby forming
 an information cascade (and therefore a herd).

\subsection{Constrained Interactive  Sensing: Individual Privacy vs Group Reputation}
The above social learning protocol can be interpreted as follows.
Agents seek to estimate an underlying state of nature
but reveal their actions by maximizing their privacy according  to  the optimization (\ref{eq:myopic}).
This leads to an information cascade and social learning stops.
In other words,
agents are interested in optimizing their own costs (such as maximizing privacy) and   ignore the information benefits their action provides
to others.

\subsubsection{Partially Observed Markov Decision Process Formulation}
We now describe an optimized social learning procedure 
that 
 delays herding.\footnote{In the restaurant problem, an obvious approach to prevent herding is as follows. If a restaurant knew that patrons choose the restaurant with the most  customers,
then the restaurant  could deliberately pay actors to sit in the restaurant so that it appears popular thereby attracting customers. The methodology in this section where herding is delayed by benevolent agents  is a different approach.}
 This approach, see \cite{Cha04} for an excellent discussion, is motivated by the following question:
How can agents assist social learning by  choosing their actions to trade off individual privacy (local costs) with
 optimizing the reputation\footnote{\cite{Mui02,GGG09} contain lucid  descriptions of 
quantitative models for trust, reputation and privacy}  of the entire social group?  

Suppose
agents seek to maximize the reputation of their social group
by  minimizing the following social welfare cost involving all agents in the social group (compared to
the myopic objective (\ref{eq:myopic})  used in standard social learning):
\beq \label{eq:pomdp}
J_\mu(\pizero) = \Ep \biggl\{\sum_{k=1}^\infty \discount^{k-1} c_{a(\pi_{k-1},y_k,\mu(\pi_{k-1}))}^\p  \eta_k 
 \biggr\}  \eeq
In (\ref{eq:pomdp}), 
$a(\pi,y,\mu(\pi)))$ denotes the decision rule that agents use to choose their actions as will be explained below. 
Also  $\discount\in [0,1)$ is an economic discount factor and $\pi_0$ denotes the initial probability (prior)  of the state $x$.
 $\Pp$ and $\Ep$ denote the probability measure and expectation
of the evolution of the observations and underlying state which are strategy dependent.

The key attribute of (\ref{eq:pomdp})  is that each agent $k$ chooses its  action according to the
privacy constrained rule  
\beq a_k = a(\pi_{k-1},y_k,\mu(\pi_{k-1})).  
\label{eq:akp}
\eeq
Here,
the policy  $$\mu: \pi_{k-1} \rightarrow \{1,2\ldots, L\}  $$
maps the available public belief to the set of $L$ privacy values.
The higher the privacy value, the less the agent reveals through its action.  
This is  in contrast  to standard social
learning  (\ref{eq:myopic}) in which the action chosen is $a(\pi,y)$, namely a myopic function of the private observation and public
belief.

The above formulation can be interpreted as follows:  Individual agents seek to maximize
their privacy according to social learning (\ref{eq:akp}) but also seek to maximize the reputation of their entire
social group (\ref{eq:pomdp}).

Determining the  policy $\mu^*$  that 
minimizes  (\ref{eq:pomdp}), and thereby maximizes the social group reputation, is equivalent to solving a   stochastic control problem that is 
called a  partially observed Markov decision process (POMDP) problem \cite{Kri11,Cas98}. 
A POMDP comprises of a noisy observed Markov chain and the dynamics of the posterior distribution (belief state) is controlled by
a policy ($\mu$ in our case).

\subsubsection{Structure of Privacy Constrained Sensing Policy}
 In general, POMDPs are computationally intractable to solve and therefore one cannot say anything useful about the structure of the optimal policy $\mu^*$.
 However, useful insight can  be obtained by considering  the following extreme case of the
  above problem.  Suppose there are two privacy values  and each agent $k$ chooses action
  $$ a_k = \begin{cases} y_k  & \text{ if } \mu(\pi_k) = 1 \text{ (no privacy)} \\
  \arg\min_a c_a^\p \pi_{k-1} & \text{ if }  \mu(\pi_k) = 2 \text{ (full privacy}).
   						   \end{cases} 
$$  
  That is, an agent either reveals its raw observation (no privacy) or chooses its action by completely neglecting its observation (full privacy).
 Once an  agent chooses the full privacy option, then all subsequent agents  choose exactly
 the same option and therefore herd - this follows since each agent's
  action reveals
  nothing about the underlying state of nature. Therefore, for this extreme example, determining the optimal policy $\mu^*(\pi)$ is equivalent to solving
  a stopping time problem:
Determine  the earliest time for agents to herd (maintain full privacy) subject to maximizing the
social group reputation.
  
  For  such a quickest herding  stopping time problem, one can say a lot about the structure of $\mu^*(\pi)$.
   Suppose the sensing system
wishes to determine if the state of nature is a specific target state (say state 1).
Then  \cite{Kri11} shows that under reasonable conditions
on the   observation distribution and supermodular conditions on the costs (\cite{MR07} discusses
supermodularity of influence in social networks),
 the dynamic programming recursion has a supermodular structure
 (see also  \cite{Lov87,Rie91,KD07,Kri12,Kri13} for related results).
This implies that the optimal policy $\mu^*$ has the following structure:
There exists a threshold curve that partitions the  belief space such that when the belief state is on one
side of the curve it is optimal for agents to reveal full observations; if the belief state is on the other side of the curve
then it is optimal to herd. Moreover, the target state~1 belongs to the region in which  it is optimal to herd.\footnote{In standard POMDPs where agents do not perform social learning, it is well known \cite{Lov87a} that the  set of beliefs
for which it is optimal to stop is convex.
Such convexity of the herding set does not hold in the current problem. But it is shown in \cite{Kri11}
 that the set of beliefs for which it is optimal 
to herd is  connected and so is the set of beliefs for which it is optimal to reveal full observations.}
This threshold structure of the optimal policy means that if individuals deploy the simple heuristic
of
\begin{quote}
``Choose increased privacy when belief is close to target state'' , \end{quote} then the group behavior
is sophisticated --  herding is delayed
 and accurate estimates of the state of nature can be obtained.

 \section{Data Incest in Online Reputation Systems} \label{sec:incest}

This section generalizes the previous section by considering social learning in a social network.
How can multiple social sensors interacting over a social network  estimate an underlying state of nature?  The state could be the position coordinates of an event \cite{SOM10} or the quality of a social parameter such as quality of a restaurant or political party.

\begin{figure}[h]
\centering
{\includegraphics[scale=0.3]{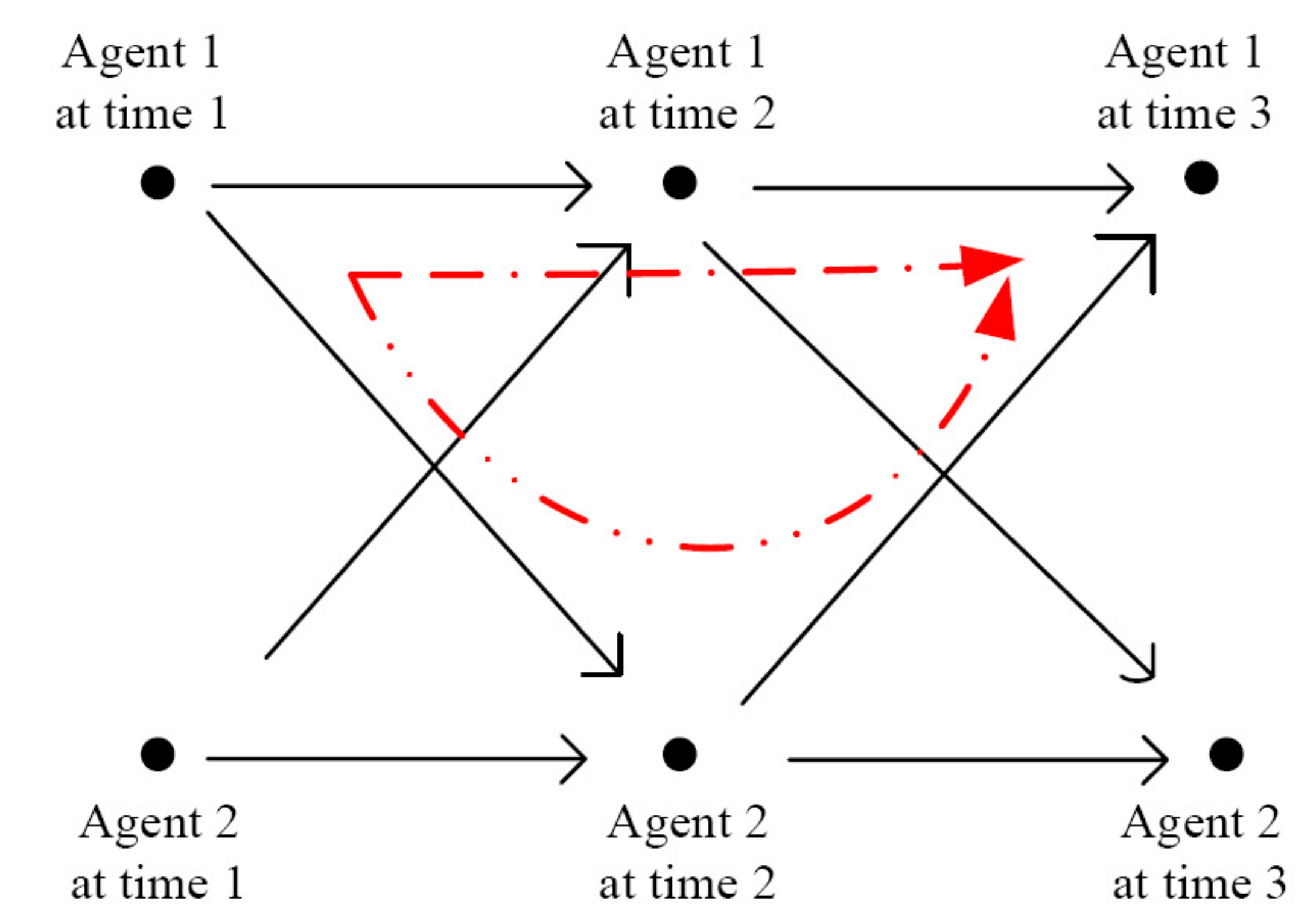}}
\caption{ Example of the information flow (communication graph) in a social network with two agents  and over three event epochs. The arrows represent exchange of information.}
\label{fig:sample}
\end{figure}

The motivation for this section can be understood in terms of the following  sensing example.
 Consider the following interactions in a multi-agent social network where  agents seek to estimate an underlying state of nature. Each agent visits a restaurant based on reviews on an online reputation website. The agent then obtains  a private measurement of the state  (e.g., the quality of food in a restaurant) in noise. After that, he reviews the restaurant on the same online reputation website.  The information exchange in the social network is modeled by a directed graph. 
 As mentioned in the introduction, data incest  \cite{KH13} arises due to  loops in the information exchange graph.
 This is illustrated in the graph of Fig.\ref{fig:sample}. Agents 1 and 2 exchange beliefs (or actions) as depicted in Fig.\ref{fig:sample}. The fact
 that there are two distinct paths between Agent 1 at time 1 and Agent 1 at time 3 (these two paths are denoted in red)  implies that 
 the information of Agent 1 at time 1 is double counted thereby leading to a data incest event.

How can data incest be removed so that agents obtain a fair (unbiased) estimate of the underlying state?
The methodology of this section  can be interpreted in terms of the recent {\em Time} article \cite{Tut13} which provides interesting rules for online reputation systems. These include: (i) review the reviewers, and  (ii) censor fake (malicious) reviewers. The data incest removal algorithm proposed in this paper can be viewed as  ``reviewing the reviews" of other agents to see if they are associated with data incest or not.

The rest of this section is organized as follows:
\begin{compactenum}
\item  Sec.\ref{sec:social} describes the social learning model that is used to mimic the behavior of agents in online reputation systems. 
The information exchange between agents in the social network is  formulated on a family of time dependent directed acyclic graphs.
\item In Sec.\ref{sec:repute}, a fair reputation protocol is presented and the criterion for achieving a fair rating is defined.
\item  Sec.\ref{sec:removal} presents an incest removal algorithm so that the online reputation system
achieves a fair rating. A necessary and sufficient condition is given on the graph structure of information exchange between agents so that a fair rating is achievable.
\end{compactenum}

\paragraph*{Related works}  Collaborative recommendation systems are reviewed and studied in \cite{AT05,KSJ09}. In
\cite{KT12},  
a model of Bayesian social learning is considered in which  agents receive private information about the state of nature and observe actions of their neighbors in a tree-based network.
 Another type of mis-information caused by influential agents (agents who heavily affect  actions of other agents in social networks) is investigated in 
\cite{AO11}. Mis-information in the context of this paper is motivated by sensor networks where the term ``data incest" is used \cite{BK07}.   Data incest also arises in  Belief Propagation (BP) algorithms \cite{Pea86, MWJ99} which are used in computer vision and error-correcting coding theory. 
BP algorithms require passing local messages over the graph (Bayesian network) at each iteration. 
For graphical models with loops, BP algorithms are only approximate due to the over-counting of local messages \cite{YFW05} which is similar to data incest in  social learning.  With the algorithms presented in this section, data incest can be mitigated from Bayesian social learning over non-tree graphs that satisfy a topological constraint.
The closest work to the current paper is \cite{KH13}. However,  in \cite{KH13}, data incest is considered in a network where agents exchange their private belief states - that is, no social learning is considered.  Simpler versions of this information exchange process and estimation were investigated in  \cite{Aum76,GP82,BV82}. 


\subsection{Information Exchange Graph in Social Network}\label{sec:social}
 Consider an online reputation system  comprised of social  sensors $\{1,2,\ldots,S\}$ that aim to estimate an underlying state of nature (a random variable). 
 Let $x \in 
 \X = \{1,2,\ldots,X\}$
 represent the state of nature (such as the quality of a hotel) with known prior distribution $\pi_0$. Let $k = 1,2,3,\ldots$ depict epochs at which events occur. These events involve taking observations, evaluating beliefs and choosing actions as  described below. The index $k$ marks the historical order of events and not necessarily absolute time. However, for simplicity, we refer to $k$ as ``time". 
 
 To model the information exchange in the social network, we will use a family of directed acyclic graphs.
 It is convenient also to reduce the coordinates of time $k$ and agent $s$  to a single integer index  $n$   which is used to represent agent $s$ at time $k$:
\begin{equation} \label{reindexing_scheme}
 n \triangleq s+ S(k-1), \quad
s \in \{1,\ldots, S\}, \; k = 1,2,3,\ldots 
\end{equation}
We  refer to $n$ as a ``node" of a time dependent information flow   graph $G_n$ that we now define.

\subsubsection{Some Graph Theoretic Definitions} 
 Let \begin{equation}\label{eq:defG} G_{n} = (V_{n}, E_{n}), \quad n  = 1,2,\ldots \end{equation} denote a sequence of time-dependent graphs of information flow in the social network until and including time $k$ where $n = s + S(k-1)$. Each vertex in $V_{n}$ represents an agent $s'$ in the social network at time $k'$ and each edge $(n',n'')$ in $E_{n}\subseteq V_{n} \times V_{n}$ shows that the information (action) of node $n'$ (agent $s'$ at time $k'$) reaches node $n''$ (agent $s''$ at time $k''$).  It is clear that the communication graph $G_n$ is  a sub-graph of $G_{n+1}$. This means that the diffusion of actions  can be modelled via a family of time-dependent directed acyclic  graphs (a directed graph with no directed cycles.

The algorithms below will involve specific columns of the adjacency matrix  transitive closure matrix  of the graph $G_n$.
The Adjacency Matrix $A_n$ of $G_n $ is an $n\times n$ matrix with elements $A_n(i,j)$  given by 
\beq \label{eq:adjacencymatrix}
A_n (i,j)=\begin{cases}
1 &\textrm{ if } (v_j,v_i)\in E \;, \\
0 &\textrm{ otherwise}
\end{cases}\;, \text{  } A_n(i,i)=0.
\eeq

The transitive closure matrix $T_n$ is the  $n\times n$ matrix 
\beq T_n =  \text{sgn}((\mathbf{I}_n-A_n)^{-1}) \label{eq:tc} \eeq
where for any matrix $M$, the matrix $\text{sgn}(M)$ has elements
$$ 
\text{sgn}(M)(i,j) = \begin{cases} 0 & \text{ if } M(i,j)=0\;, \\
1  & \text{ if } M(i,j) \neq 0. \end{cases}
$$
Note that $A_n(i,j) = 1$ if there is a single hop path between nodes $i$ and $j$, In comparison,
$T_n(i,j) = 1$ if there exists a path (possible multi-hop) between node $i$ and $j$.

The information reaching node $n$ depends on the information flow graph  $G_n$.
The following two sets will be used  to specify the incest removal algorithms below:
\begin{align}
\history_n &= \{m : A_n(m,n) = 1 \}  \label{eq:history}  \\
\full_n &= \{m : T_n(m,n) = 1 \} . \label{eq:full}  
\end{align}
Thus $\history_n$ denotes the set of previous nodes $m$ that communicate with node $n$ in a single-hop.
In comparison, $\full_n$
  denotes the set of previous nodes $m$ whose information eventually arrive at node $n$. Thus $\full_n$  contains all possible  multi-hop connections by which information from a node $m$
 eventually reaches node $n$. 

\subsubsection{Example} To illustrate the above notation 
consider a social network consisting of $S=2 $ two groups with the following information flow graph for three  time points $k=1,2,3$.
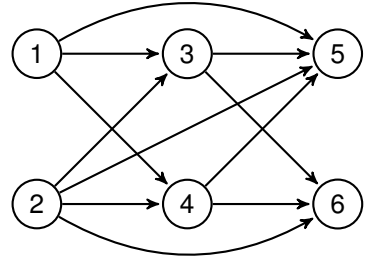
\begin{figure}[h]
\centering
\begin{tikzpicture}[->,>=stealth',shorten >=1pt,auto,node distance=2cm,
  thick,main node/.style={circle,fill=white!12,draw,font=\sffamily}]

  \node[main node] (1) {1};
  \node[main node] (2) [below  of=1] {2};
  \node[main node] (3) [right of=1] {3};
  \node[main node] (4) [below  of=3] {4};
  \node[main node] (5) [right of=3] {5};
  \node[main node] (6) [right of=4] {6};

  \path[every node/.style={font=\sffamily\small}]
    (1) edge node  {} (4)
        edge [bend left] node[left] {} (5)
        edge node {} (3)
    (2) edge node {} (3)
        edge node {} (4)
        edge  node {} (5)
        edge [bend right] node[left] {} (6)
    (3) edge node  {} (5)
        edge node  {} (6)
    (4) edge node  {} (5)
        edge node {} (6);
\end{tikzpicture}
\caption{Example of an  information flow network with $S=2$  two agents, namely   $s\in \{1,2\}$ and  time points $k=1,2,3$.  Circles represent the nodes indexed by $n= s + S(k-1)$
in the social network and each edge depicts a communication link between two nodes.}
\label{sample}
\end{figure}

  Fig.\ref{sample} shows the nodes  $n =1,2,\ldots,6$ where $n= s + 2(k-1)$. 
  
  Note that in this example, as is apparent from Fig.\ref{fig:sample}, each node remembers all its previous actions.
  The information flow is characterized by 
the family of directed acyclic graphs  $\{G_1,G_2,G_3,G_4,G_5,G_6\}$  with adjacency matrices  
\\
$A_1 = \begin{bmatrix}
0
\end{bmatrix}$,
$A_2 = \begin{bmatrix}
0 & 0 \\
0 & 0
\end{bmatrix}$,
$A_3 = \begin{bmatrix}
0 & 0 & 1 \\
0 & 0 & 1 \\
0 & 0 & 0
\end{bmatrix}$,
$$A_4 = \begin{bmatrix}
0 & 0 & 1 & 1 \\
0 & 0 & 1 & 1 \\
0 & 0 & 0 & 0 \\
0 & 0 & 0 & 0
\end{bmatrix}, \;
A_5 = \begin{bmatrix}
0 & 0 & 1 & 1 & 1 \\
0 & 0 & 1 & 1 & 1 \\
0 & 0 & 0 & 0 & 1 \\
0 & 0 & 0 & 0 & 1 \\
0 & 0 & 0 & 0 & 0
\end{bmatrix}.$$
Since nodes 1 and 2 do not communicate,
clearly $A_1$ and $A_2$ are zero matrices. Nodes 1 and  3 communicate as do nodes 2 and  3, hence  $A_3$ has two
ones, etc.
Finally from (\ref{eq:history}) and (\ref{eq:full}),  
$$\history_5=\{1,2,3,4\}, \quad \full_5 = \{1,2,3,4\} $$
where $\history_5$ denotes all one hop links to node 5 while $\full_5$ denotes all multihop links to node 5.

Note that $A_n$ is always the upper left  $n\times n$ submatrix of $A_{n+1}$. Also
 due to causality with respect to the time index $k$, the adjacency matrices are  always upper triangular.

\subsection{Fair Online Reputation System} \label{sec:repute}

\subsubsection{Protocol for Fair Online Reputation System}
The procedure summarized in Protocol \ref{protocol:socialcons}  aims to evaluate a fair reputation that uses
 social learning over a social network by eliminating incest.

\begin{algorithm}\floatname{algorithm}{Protocol}
(i) {\em  Information from Social Network}: 
\begin{compactenum}
\item {\em Recommendation from friends}: Node $n$ receives past actions  $\{a_m, m \in \history_n\}$ from previous nodes  $m \in \history_n$ in the social network. $\history_n$ is defined in (\ref{eq:history}).
\item  {\em Automated Recommender System}: For these past actions $\{a_m, m \in \history_n\}$,   the network administrator 
has already computed the public beliefs $(\pi_m, m \in \history_n)$ using Step (v) below.\\
The automated recommender system    fuses  public beliefs 
$(\pi_m, m \in \history_n)$, into  the single   recommendation  belief  $\belief_{n-}$ as
\beq  \belief_{n-} = \mathcal{A} (\pi_m, m \in \history_n)  \label{eq:fuse}\eeq
The  fusion algorithm $\mathcal{A}$ will be designed below.
\end{compactenum}
(ii) {\em  Observation}: Node $n$  records   private observation  $y_n$ from  distribution $B_{iy} = P(y|x=i)$, $i \in \X$.\\
(iii) {\em Private Belief}:  Node $n$ then  uses $y_n$ and public belief $\belief_{n-}$ to update its private belief via Bayes formula as
\beq   \priv_n = 
\frac{B_{y_n}  \belief_{n-}}{ \mathbf{1}_X^\p B_y  \belief_{n-}}  \label{eq:privi} \eeq
\\
(iv) {\em Myopic Action}: Node $n$ takes action
$$a_n = \arg\min_a c_a^\p \eta_n$$ and inputs its action to the online reputation system.\\
(v) {\em Public Belief Update by Network Administrator}: Based on action $a_n$,
the network administrator (automated algorithm)  computes the public belief $\pi_n$ using  the social learning
filter (\ref{eq:piupdate}) with $P = I$.
\caption{Incest Removal for Social Learning  in an Online Reputation System}\label{protocol:socialcons}
\end{algorithm}

{\bf Aim}:
Our aim is to design  algorithm $\mathcal{A}$ in the automated recommender system (\ref{eq:fuse}) of Protocol \ref{protocol:socialcons} so that the following requirement is met:
\begin{align}
 \belief_{n-}(i) & =  \tbelief_{n-}(i) , \quad i \in \X \nonumber \\
 \text{ where } & \tbelief_{n-}(i) = P(x = i | \{a_m, m \in \full_n\}).   \label{eq:aims}\end{align}
We  call $\tbelief_{n-}$ in (\ref{eq:aims})  the {\em true}  or {\em fair online rating} available to
  node $n$
 since 
  $\full_n = \{m: T_n(m,n) = 1\}$ defined in (\ref{eq:full}) denotes all  information (multi-hop links) available  to node $n$. By definition
  $\tbelief_{n-}$ is incest free since it is the desired conditional probability that we want.
If algorithm $\mathcal{A}$ is designed so that $\belief_{n-}(i)$ satisfies (\ref{eq:aims}),  then  the computation (\ref{eq:privi}) and Step (v) yield
\begin{align*}   \eta_n(i) &=  P(x = i | \{ a_m, m \in \full_n\},  y_n ), \quad i \in \X,  \\
\pi_n(i)  &= P(x = i | \{a_m, m \in \full_n\},  a_n ),\quad i \in \X, \end{align*}
which are, respectively,   the correct private belief for node $n$ and the correct after-action public belief.

\subsubsection {Discussion of Protocol \ref{protocol:socialcons}} \mbox{} \\
(i) {\em Data Incest}: It is important to note that without careful design of  algorithm $\mathcal{A}$, 
due to  loops in the dependencies of actions on previous actions, the public rating $\belief_{n-}$ computed using (\ref{eq:fuse}) can be substantially different
from the fair online rating $\tbelief_{n-}$ of (\ref{eq:aims}).
As a result,  $ \eta_n $ computed via (\ref{eq:privi}) will not be the correct private belief and 
  incest will propagate in the network.  In other words,  $\eta_n$, $\belief_{n-}$ and $\belief_n$ are defined purely in terms of their computational expressions in Protocol \ref{protocol:socialcons} -- at this stage
  they are not necessarily the desired  conditional probabilities, unless algorithm $\mathcal{A}$ is designed to remove incest.
  
  Note that instead of (\ref{eq:fuse}), node $n$ could  naively (and incorrectly)  assume that the public beliefs $\belief_m, m\in \history_n$ that it received are 
  independent. 
It would then   
  fuse  these public beliefs as
  \beq  \belief_n{-} = \frac{\sum_{m\in \history_n} \belief_m }{\mathbf{1}_X^\p \sum_{m\in \history_n} \belief_m}. \label{eq:dataincest}\eeq
  This, of course, would result in data incest.
\\  
(ii) {\em How much does an individual remember?}: The above protocol has the flexibility of modelling  cases where 
either each node remembers some (or all) of its past actions or none of its past actions. This facilitates modelling cases in which
people forget most of the past except for specific highlights. \\
(iii) {\em Automated Recommender System}: Steps (i) and (v)  of Protocol \ref{protocol:socialcons} can be combined into an automated recommender system  that maps previous actions of agents in the social group
to a single recommendation (rating) $\belief_{n-}$ of (\ref{eq:fuse}). This recommender system can operate completely opaquely to the actual user (node $n$). Node $n$ simply
uses the automated rating $\belief_{n-}$ as the current best available rating from the reputation system.
\\
(iii) {\em Social Influence. Informational Message vs Social Message}: In Protocol \ref{protocol:socialcons}, it is important that each individual $n$ deploys
Algorithm $\mathcal{A}$  to fuse  the  beliefs $\{\belief_m, m \in \history_n\}$; otherwise incest can propagate.
Here,  $ \history_n$ can be viewed as the  ``social message'', i.e., personal friends of node $n$ since they directly communicate to node $n$ while the associated beliefs can be viewed as the
``informational message''.
The social message  from personal friends exerts a  large social influence -- it provides significant incentive (peer pressure) for individual $n$ to comply with
Protocol \ref{protocol:socialcons} and thereby prevent incest.
 Indeed, a remarkable recent study described in \cite{BFJ12} shows that  social messages (votes)
from known friends has significantly more influence on an individual than the information in the messages themselves. This study includes comparison of information messages and social messages on Facebook and their direct
effect on voting behavior.
 To quote \cite{BFJ12}, ``The effect of social transmission on real-world voting
was greater than the direct effect of the messages themselves..."  

(iv) {\em Agent Reputation}:
The  cost function minimization in Step (iv) can be interpreted in terms of the reputation of agents in online reputation systems. If an agent continues to write bad reviews for high quality
restaurants on Yelp, his reputation becomes lower among the users. Consequently, other people ignore reviews of that (low-reputation) agent in evaluating their opinion about the social unit under study (restaurant). Therefore, agents  minimize the penalty of writing inaccurate reviews (or equivalently increase their reputations) by choosing proper actions. 
\\
(v) {\em Think and act}: Steps (ii), (iii)  (iv) and (v)  of Protocol \ref{protocol:socialcons} constitute standard social learning as described in Sec.\ref{sec:herd}.
The key difference with standard social learning is  Steps (i) performed by the network administrator.
Agents  receive public beliefs from the social network with arbitrary random delays.
These delays reflect the time an agent takes between reading the publicly available reputation and making its decision. It is typical behavior of people to
 read  published ratings multiple times and then think for an arbitrary amount of time before acting. 
 
 \subsection{Incest Removal Algorithm in  Online Reputation System} \label{sec:removal}
 Below we design algorithm $\mathcal{A}$ in Protocol~\ref{protocol:socialcons} so that it yields the  fair  public rating $\tbelief_{n-}$ of  (\ref{eq:aims}).

\subsubsection{Fair Rating Algorithm}
 It is convenient to work with the logarithm of the un-normalized belief\footnote{The un-normalized belief proportional to $\belief_n(i)$ is the numerator of the social learning filter (\ref{eq:piupdate}).
The corresponding un-normalized fair rating corresponding to $\tbelief_{n-}(i) $ is the joint distribution $P(x=i, \{a_m, m \in \full_n\})$.
By taking logarithm of the un-normalized belief, Bayes formula merely becomes the sum of the log likelihood and log prior. This allows
 us to devise a data incest  removal algorithm based on linear combinations of the log beliefs.};  accordingly define
$$  \lbelief_n(i) \propto \log \belief_n(i), \quad \lbelief_{n-}(i) \propto  \log \belief_{n-}(i), \quad i \in \X.$$

 The following theorem shows that the logarithm of the fair rating $\tbelief_{n-}$ defined in (\ref{eq:aims}) can be obtained as linear weighted combination of the logarithms of previous public beliefs.

  \begin{theorem}[Fair Rating Algorithm] \label{thm:socialincestfilter} Consider the online reputation system running  Protocol \ref{protocol:socialcons}.
Suppose the  following algorithm 
$\mathcal{A}(\lbelief_m, m \in \history_n)$ is
implemented in (\ref{eq:fuse})  of  Protocol~\ref{protocol:socialcons} by the network administrator:
\begin{align}\label{eq:socialconstraintestimate}
\lbelief_{n-}(i)  &=   w_n^\p \, \lbelief_{1:n-1}(i) 
 \nonumber \\
\text{ where } &\;  w_n =  T_{n-1}^{-1}  t_n.
\end{align}
Then  $\lbelief_{n-}(i) \propto \log \tbelief_{n-}(i)$. That is, algorithm $\mathcal{A}$ computes the fair  rating $\log \tbelief_{n-}(i)$ defined in (\ref{eq:aims}). \\
In (\ref{eq:socialconstraintestimate}),  $w_n$ is an  $n-1$ dimensional weight vector.
Recall  that
$t_n$ denotes the first $n-1$ elements of the  $n$th column of transitive closure matrix $T_n$. \qed
\end{theorem}

Theorem  \ref{thm:socialincestfilter}  says that  the fair rating  $\tbelief_{n-}$ can be expressed as a linear function of the action log-likelihoods
 in terms of the transitive closure matrix $T_n$ of graph $G_n$. This is 
intuitive since $\tbelief_{n-}$ can be viewed as the sum of information collected by the nodes such that there are paths between all these nodes and $n$.

\subsubsection{Achievability of Fair Rating by Protocol \ref{protocol:socialcons}}
We are not quite done!  
\begin{compactenum}
\item On the one hand,  algorithm $\mathcal{A}$  at node $n$ specified by  (\ref{eq:fuse}) has access only to beliefs
$\lbelief_m, m \in \history_n$ -- equivalently  it  has access only to beliefs from previous nodes specified by $A_n(:,n)$ which denotes the  last column of the adjacency matrix $A_n$.
\item 
On the other hand,  to provide incest free estimates, algorithm $\mathcal{A}$ specified in (\ref{eq:socialconstraintestimate})  requires  all previous beliefs $l_{1:n-1}(i)$ that are specified by the non-zero elements of the vector $ w_n$.
\end{compactenum}
The only way to reconcile points 1 and 2 is  to  ensure  that  $A_n(j,n) = 0$ implies $w_n(j) = 0$ for $j=1,\ldots, n-1$. The condition means that the single hop 
past estimates $\lbelief_m, m \in \history_n$
available at node $n$ according to  (\ref{eq:fuse}) in Protocol \ref{protocol:socialcons}  provide all the information required to compute 
$w_n^\p \, \lbelief_{1:n-1}$ in (\ref{eq:socialconstraintestimate}).
  This is a condition on the information flow graph $G_n$.
We formalize this condition in 
the following theorem.

\begin{theorem}[Achievability of Fair Rating]\label{thm:sufficient}
Consider the fair rating algorithm specified by (\ref{eq:socialconstraintestimate}). For Protocol \ref{protocol:socialcons}  with available information $(\belief_m, m \in \history_n)$ to achieve the estimates $\lbelief_{n-}$ of algorithm (\ref{eq:socialconstraintestimate}),
a necessary and sufficient condition on the information flow graph $G_n$ is \beq \label{constraintnetwork}
A_n(j,n)=0   \implies w_n(j)= 0.
\eeq
Therefore for Protocol  \ref{protocol:socialcons} to generate incest free estimates for nodes $n=1,2,\ldots$, condition  (\ref{constraintnetwork}) needs to hold for each $n$.
(Recall 
  $w_n$ is specified in (\ref{eq:socialconstraintestimate}).)
\qed
\end{theorem}

Note that the constraint (\ref{constraintnetwork}) is purely in terms of the adjacency matrix $A_n$, since the transitive closure matrix (\ref{eq:tc}) is a function of the adjacency matrix.

\noindent{\em Summary}: Algorithm (\ref{eq:socialconstraintestimate}) together with the condition (\ref{constraintnetwork}) ensure that incest free estimates are generated
by Protocol \ref{protocol:socialcons}.

\subsubsection{Illustrative Example (continued)} Let us continue with the example of Fig.\ref{fig:sample} where we already specified the adjacency matrices of the graphs
$G_1$, $G_2$, $G_3$, $G_4$ and $G_5$.
Using (\ref{eq:tc}), the transitive closure matrices $T_n$  obtained from the adjacency matrices are given by:\\
\noindent{\small
$T_1 = \begin{bmatrix}
1
\end{bmatrix},
T_2 = \begin{bmatrix}
1 & 0 \\
0 & 1
\end{bmatrix},
T_3 = \begin{bmatrix}
1 & 0 & 1 \\
0 & 1 & 1  \\
0 & 0 & 1
\end{bmatrix}$, \\ $
T_4 = \begin{bmatrix}
1 & 0 & 1 & 1  \\
0 & 1 & 1 & 1 \\
0 & 0 & 1 & 0 \\
0 & 0 & 0 & 1
\end{bmatrix},
T_5 = \begin{bmatrix}
1 & 0 & 1 & 1 & 1 \\
0 & 1 & 1 & 1 & 1 \\
0 & 0 & 1 & 0 & 1 \\
0 & 0 & 0 & 1 & 1 \\
0 & 0 & 0 & 0 & 1
\end{bmatrix}$.
}\\
Note that $T_n(i,j) $ is non-zero only for $i\geq j$ due to causality since information sent by a social group can only arrive at another social group at a later time instant. The weight vectors are then
obtained from  (\ref{eq:socialconstraintestimate}) as
$\\
w_2 = \begin{bmatrix}0\end{bmatrix},\;\\
w_3 = \begin{bmatrix}1 & 1\end{bmatrix}^\p,\;\\
w_4 = \begin{bmatrix}1 & 1 & 0\end{bmatrix}^\p,\;\\
w_5 = \begin{bmatrix}-1 & -1 & 1 & 1\end{bmatrix}^\p.\\
$
Let us examine these weight vectors. $w_2$ means that node $2$ does not use the estimate from node $1$. This formula is consistent with the constraint information flow because estimate from node $1$ is not available to node $2$; see Fig.\ref{sample}.
$w_3$ means that node $3$ uses estimates from node $1$ and $2$; $w_4$ means
that  node $4$  uses estimates only from node $1$ and node $2$. The estimate from node $3$ is not available at node 4. As shown in Fig.\ref{sample}, the mis-information propagation occurs at node $5$. The vector $w_5$ says that node 5 adds estimates from nodes $3$ and $4$ and removes estimates from nodes $1$ and $2$ to avoid double counting of these estimates already integrated into estimates from node $3$ and $4$. Indeed, using the algorithm (\ref{eq:socialconstraintestimate}), incest is completely prevented in this example. 

Let us now illustrate an example in which exact incest removal is impossible.
Consider the information flow graph of  Fig.\ref{sample} but with the edge between node 2 and node 5  deleted. Then $A_5(2,5) = 0$ while $w_5(2) \neq 0$, and therefore the condition (\ref{constraintnetwork}) does not hold. Hence exact incest  removal is not possible for this case.

\subsection{Summary}
In this section,
we have outlined a controlled sensing problem over a social network in which the administrator controls (removes) data incest
and thereby maintains an unbiased (fair) online reputation system.  The state of nature could be geographical coordinates of an event (in a target localization problem) or quality of a social unit (in an online reputation system).  As discussed above, data incest arises  due to the recursive nature of Bayesian estimation and non-determinism in the timing of 
the  sensing by individuals. Details of proofs, extensions and further numerical studies are presented in~\cite{KH13,HK13}.

\section{Interactive Sensing for  Quickest  Change  Detection}  \label{sec:socialc}

In this section we 
consider interacting social sensors in the context of   detecting a change in the underlying state of nature.
Suppose a multi-agent system  performs social learning and makes local decisions as described in Sec.\ref{sec:classicalsocial}. Given the public beliefs from the social learning protocol, how can quickest  change detection be achieved?
In other words, how can a global decision maker use the local decisions from individual agents to decide when a change has occurred?
It is shown below that making a global decision (change or no change) based on local decisions of individual agents has an unusual structure resulting in a non-convex stopping set.

A typical application of such social sensors  arises in the  measurement of the adoption of a new product using a micro-blogging platform like Twitter. 
The adoption of the technology diffuses through the market but its effects can only be observed through the tweets of select individuals of the population.
 These selected individuals act as  sensors for estimating the diffusion. They interact and learn from the decisions (tweeted sentiments) of  other members and
 therefore perform
 social learning.  Suppose the state of nature suddenly changes due to a sudden market shock or presence of a new competitor.
 The goal for a market analyst or product manufacturer is to detect this change as quickly as possible by minimizing a cost function that involves the sum of the  false alarm and decision delay.

 \paragraph*{Related works}
 \cite{Pin06,Pin08} model diffusion in networks over a random graph with arbitrary degree distribution. The resulting diffusion is approximated using deterministic dynamics via a mean field approach
 \cite{BW03}. In the seminal paper \cite{EP06}, a sensing system for complex social systems is presented with data collected
 from cell phones.  This data is used in \cite{EP06} to recognize social patterns, identify socially significant locations and infer relationships.
 In \cite{SOM10}, people using a microblogging service such as Twitter are considered as  sensors.
 In particular, \cite{SOM10} considers each  Twitter user as a sensor and uses a particle filtering  algorithm to estimate the  centre of 
 earthquakes and trajectories of typhoons. As pointed out in \cite{SOM10}, an important characteristic of microblogging services such as Twitter is that they
 provide  real-time  sensing -- Twitter users tweet several times a day; whereas  standard blog users  update information  once every several days.

Apart from the above applications in real time  sensing,  change detection in social learning 
also arises in mathematical finance models.
For example, in agent based models  for the microstructure of asset prices in high frequency trading in financial systems \cite{AS08}, the
 state denotes the underlying
asset value that changes at a random time $\tau^0$. 
Agents observe local individual decisions of previous agents via an order book, combine these observed decisions with their noisy private signals about the asset, selfishly optimize their expected local utilities, and then make their own individual decisions (whether to buy, sell or do nothing).
  The market evolves through the orders of trading agents. 
 Given this order book information, the goal of the market maker (global decision maker) is to achieve quickest change point detection when a shock occurs to the value of the asset~\cite{KA12}.

\subsection{Classical Quickest Detection}
The classical Bayesian quickest time detection problem \cite{PH08} is as follows: 
An underlying discrete-time state process $x$ jump-changes at a geometrically distributed random time $\tau^0$.
Consider a sequence of discrete time random measurements $\{y_k,k \geq 1\}$, such that 
 conditioned on the event $\{\tau^0 = t\}$, $y_k$, $k \leq t$  are independent and identically distributed (i.i.d.) random variables with distribution 
$B_1$ and $y_k, k >t$ are i.i.d. random variables with distribution $B_2$.
The quickest  detection problem involves detecting the change time $\tau^0$ with minimal cost. That is,
at each time $k=1,2,\ldots$, a decision $u_k \in \{ 1 \text{ (stop and announce change)}, 2 \text{ (continue)} \}$ needs to be made to optimize a tradeoff
between false alarm frequency and linear delay penalty.

To formalize this setup, let $\tp = \begin{bmatrix} 1 & 0 \\ 1-\tp_{22} & \tp_{22} \end{bmatrix}$ denote the transition matrix of  a two state Markov chain $x$
in which state 1 is absorbing. Then it is easily seen that the geometrically distributed change time $\tau^0$ is equivalent to  the time
at which the Markov chain enters state 1. That is $\tau^0 = \min \{k: x_k = 1\}$ and $\E\{\tau^0\} = 1/(1-\tp_{22})$.
Let $\tau$ be the time at which the decision  $u_k = 1$ (announce change) is
taken. The goal of quickest time detection is to minimize the  Kolmogorov--Shiryaev
criterion for detection of  a disorder \cite{Shi63}:
\beq J_\mu(\pizero) =   d \,\Ep\{(\tau - \tau^0)^+\} +  f \Pp(\tau < \tau^0) .
\label{eq:ksd} \eeq
Here   $x^+ = x$ if $x\geq  0$ and $0$ otherwise. The non-negative constants $d$ and $f$ denote  the delay  and false alarm penalties, respectively.
So waiting too long to announce a change incurs a delay penalty $d$ at each time instant after the system has changed, while declaring
a change before it happens, incurs a false alarm penalty $f$.
In (\ref{eq:ksd})  $\mu$ denotes the  strategy of  the decision maker.  $\Pp$ and $\Ep$ are the probability measure and expectation
of the evolution of the observations and Markov state which are strategy dependent. $\pi_0$ denotes the initial distribution of the Markov chain $x$. 

In  classical quickest detection, the  decision policy $\mu$  is a function of the two-dimensional
belief  state (posterior probability mass function)  $\pi_k(i) = P(x_k = i | y_1,\ldots,y_k,u_1,\ldots,u_{k-1})$,
$i=1,2$, with  $\pi_k(1)+\pi_k(2) =1$.
So  it suffices to consider one element, say $\pi_k(2)$,  of this
probability mass function. Classical quickest  change detection (see for example  \cite{PH08}) says that the policy $\mu^*(\pi)$ which
optimizes (\ref{eq:ksd}) has the following  threshold structure:
There exists a threshold point $\pi^* \in [0,1]$ such that  
\beq \mu^*(\pi_k) = \begin{cases} 2  \text{ (continue) } & \text{ if }
\pi_k(2) \geq \pi^* \\  1  \text{ (announce change)  } &  \text{ if } \pi_k(2) < \pi^*.
\end{cases} \label{eq:onedim}
\eeq  

\subsection{Multi-agent Quickest Detection Problem}
With the above classical formulation in mind, consider now the following multi-agent quickest change detection problem.
Suppose that a multi-agent system  performs
social learning  to estimate an underlying  state according to the social learning protocol of Sec.\ref{sec:herd}. 
 That is,
each agent acts once in a predetermined sequential order indexed by $k=1,2,\ldots$ (Equivalently, as pointed out in the discussion
in Sec.\ref{sec:herd},  a finite number of agents act repeatedly in some pre-defined order and each
 agent chooses its local decision using the current public belief.)
  Given these local decisions (or equivalently the public belief),  the goal of the global decision maker is to minimize the quickest detection objective   (\ref{eq:ksd}).
The problem now is a 
 non-trivial generalization of classical quickest detection. The posterior $\pi$ is now the public belief given by the social learning filter (\ref{eq:piupdate})
 instead of a standard Bayesian filter.
There is now  interaction between  the local and global decision makers. The local decision $a_k$ from the social learning protocol determines the public belief state $\pi_k$ via the social learning filter (\ref{eq:piupdate}), which determines
the global decision (stop or continue), which determines the local decision at the next time instant, and so on.

The global decision maker's policy $\mu^*:\pi \rightarrow \{1,2\}$ that optimizes the quickest detection objective (\ref{eq:ksd}) and
the cost $J_{\mu^*}(\pi_0)$ of this optimal policy are the solution
of 
 ``Bellman's dynamic programming  equation''  
\begin{align} \label{eq:dp_alg}
\mu^*(\pi)&= \arg\min\{ f \pi(2), \; d(1-\pi(2)) \nonumber
\\ & + \sum_{a \in \A}  V\left( T(\pi ,a) \right) \sigp(\pi,a)\} , \quad J_{\mu^*}(\pi_0) = V(\pi_0) \nonumber \\
 V(\pi) &= \min \{  f \pi(2), \; d(1-\pi(2)) \nonumber \\ 
& \hspace{1.5cm} +  \sum_{a \in \A}  V\left( T(\pi,a) \right) \sigp(\pi,a)\}.
\end{align}
Here  $T(\pi,a)$  and $\sigma(\pi,a)$ are given by the social learning filter
(\ref{eq:piupdate}) - recall   that $a$ denotes the local decision.
$V(\pi)$ is called the ``value function" -- it is the cost incurred by the optimal policy when the initial belief state (prior) is $\pi$.
As will be shown the numerical example below, the optimal policy $\mu^*(\pi)$ has a very different structure compared to classical quickest detection.

\subsection{Numerical Example} \label{sec:numerical}
We  now illustrate the unusual multi-threshold property of the global decision maker's optimal policy $\mu^*(\pi)$ in multi-agent quickest detection
with social learning.
Consider the social learning model of Sec.\ref{sec:herd}  with the following parameters:
The underlying state is a 2-state Markov chain $x$ with state space $\X= \{1,2\}$ and transition probability matrix $ \tp =  \begin{bmatrix} 1 &  0 \\  0.05 & 0.95 \end{bmatrix}$. So the change time $\tau^0$  (i.e., the time the Markov chain jumps  from state 2 into absorbing state 1) is geometrically distributed with $E\{\tau^0\} = 1/0.05 = 20$.

\begin{figure}\centering
\mbox{\subfigure[Optimal global decision policy $\mu^*(\pi)$]
{\includegraphics[scale=0.22]{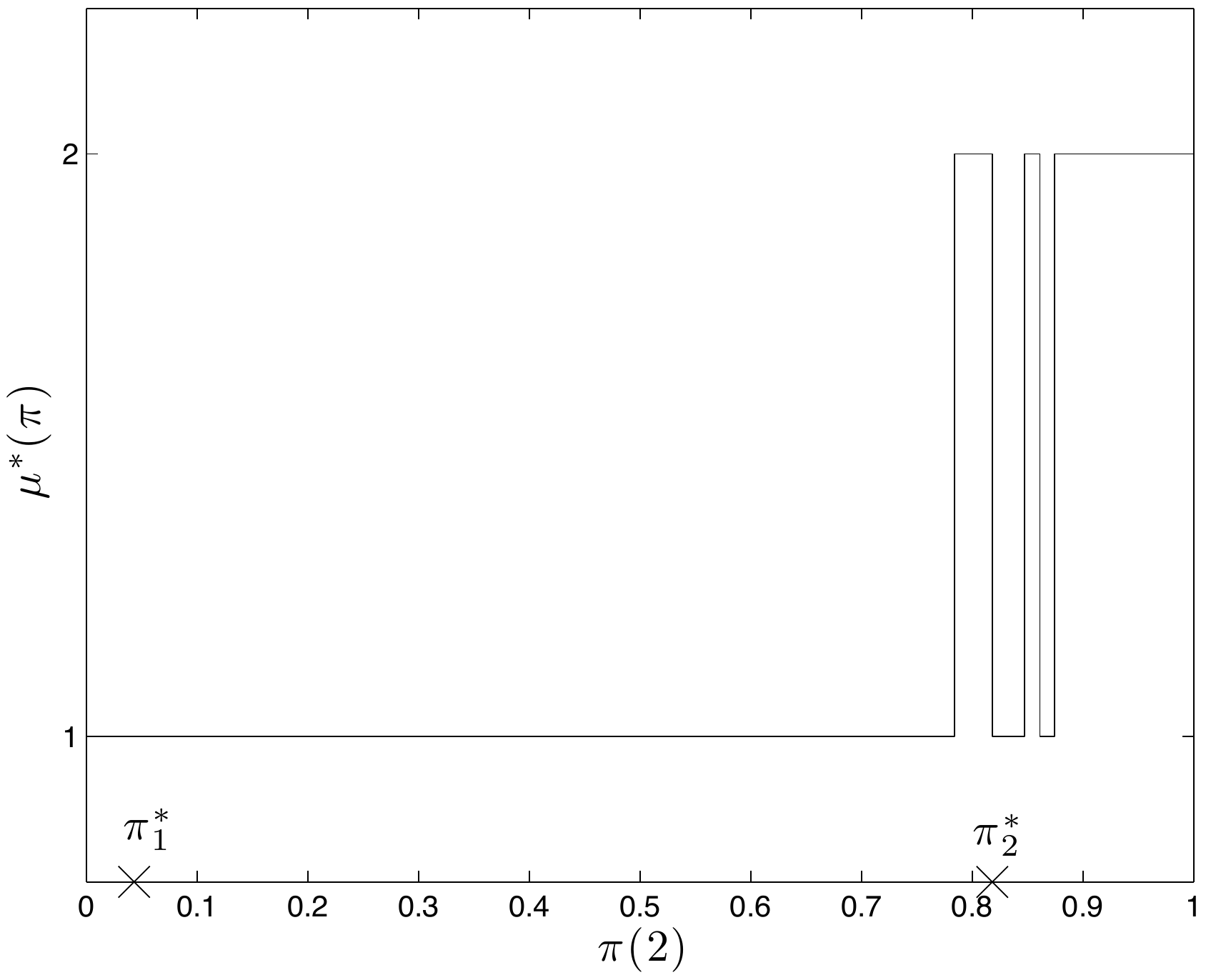}} \quad
\subfigure[Value function $V(\pi)$ for  global decision policy]
{\includegraphics[scale=0.22]{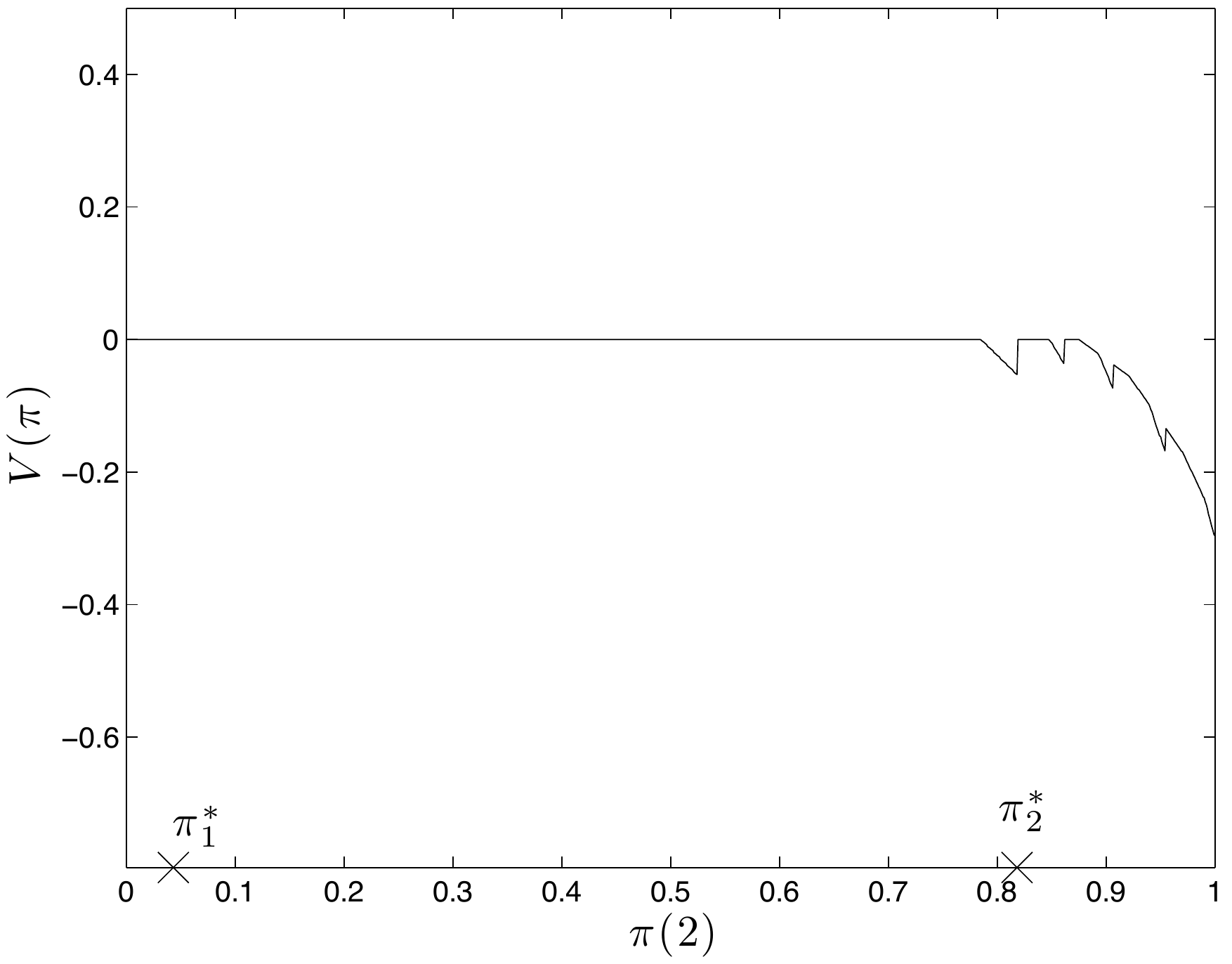}}   }
\caption{Optimal global decision policy for social learning based quickest time change detection 
for geometric distributed  change time.  The parameters are specified in Sec.\ref{sec:numerical}.  The optimal policy $\mu^*(\pi)\in \{1 \text{ (announce change) }, 2 \text{ (continue) }\}$ is characterized by a triple threshold -- that is, it switches from 1 to 2 three times as the posterior $\pi(2)$ increases. The value function $V(\pi)$ is non-concave and discontinuous in $\pi$.  As explained in the text, for $\pi(2) \in [0,\pi_1^*]$, all agents herd, while for $\pi(2) \in [\pi_2^*,1]$ individual agents herd (see
definitions in Sec.\ref{sec:cascade}).}
\label{fig:redgreen}
\end{figure}

{\em Social Learning Parameters}: Individual agents observe the Markov chain $x$ in noise with the observation symbol set $\Y=\{1,2\}$.  Suppose the observation
likelihood matrix with elements  $B_{iy} = P(y_k=y | x_k = i)$ is
 $B = \begin{bmatrix}  0.9  & 0.1  \\  0.1 & 0.9 \end{bmatrix}$.
 Agents can choose their local actions $a$ from the action set $\A=\{1 ,2 \}$.
The state dependent cost matrix of these actions is  $c= (c(i,a), i\in X, a\in \A) = \begin{bmatrix} 
4.57 & 5.57 \\ 2.57 & 0
 \end{bmatrix}$.
 Agents perform social learning with the above parameters.
  The intervals $ [0,\pi_1^*]$ and 
$ [\pi_2^*,1] $ in Fig.\ref{fig:redgreen}(a) are regions where the optimal local actions taken by agents are independent of their observations.
For $\pi(2) \in  [\pi_2^*,1] $, the optimal local action is 2 and  for  $\pi(2) \in [0,\pi_1^*]$, the optimal local action is 1.
So individual agents herd for belief states in these intervals (see the definition in Sec.\ref{sec:cascade})
and the local actions 
 do not yield any information about the underlying state.
Moreover, the interval  $[0,\pi_1^*]$ depicts a region where all agents herd (again see  the definition in Sec.\ref{sec:cascade}),  meaning that once the belief state is in this region, it remains so 
indefinitely and all agents choose the same local action 1.\footnote{Note that even if the agent $k$ herds so that its  action $a_k$
provides no information about its private observation $y_k$, the public belief still evolves according
to the predictor $\pi_{k+1} = \tp^\p \pi_k$. So an information cascade does not occur in this example.}

{\em Global Decision Making}: Based on the local actions of the agents performing social learning, the global decision maker needs to perform quickest  change detection.
The global decision maker uses the 
  delay penalty $d=1.05$ and false alarm penalty  $f=3$ in the objective function (\ref{eq:ksd}).
The optimal policy $\mu^*(\pi)$ of the global decision maker where $\pi = [1-\pi(2), \pi(2)]^\p$ is plotted versus $\pi(2)$  in Fig.\ref{fig:redgreen}(a).
Note $\pi(2) = 1$ means that with certainty no change has occurred, while $\pi(2) = 0$ means with certainty a change has occurred.
The policy $\mu^*(\pi)$
was computed by constructing
a uniform grid of 1000 points for $\pi(2)\in [0,1]$ and then implementing the dynamic programming equation
(\ref{eq:dp_alg}) via a fixed point value  iteration algorithm
for 200 iterations.
The horizontal axis $\pi(2)$ is the posterior probability
of no change.
The vertical axis denotes the optimal decision:   $u=1$ denotes stop and declare change, while
 $u=2$ denotes continue.
 
 The most remarkable feature of Fig.\ref{fig:redgreen}(a) is  the multi-threshold behavior of  the 
global decision maker's optimal policy $\mu^*(\pi)$. Recall $\pi(2)$ depicts the posterior probability of no change.
 So consider the region where  $\mu^*(\pi)  = 2$ and sandwiched between
 two regions where $\mu^*(\pi) = 1$.  Then as  $\pi(2)$ (posterior probability of  no change) increases, the optimal policy
  switches from $\mu^*(\pi) = 2$ to $\mu^*(\pi) = 1$. In other words, the optimal global decision policy ``changes its mind" -- 
  it 
  switches from no change to  change as  the posterior probability of a change decreases!
 Thus, the global decision (stop or continue) is a non-monotone function of the posterior probability obtained from local decisions.

Fig.\ref{fig:redgreen}(b) shows the associated value function obtained via stochastic dynamic programming (\ref{eq:dp_alg}). Recall that
$V(\pi)$  is the cost
incurred by the optimal policy with initial belief state $\pi$.
Unlike standard sequential detection problems in which the value function is concave, the figure shows 
that the value function is non-concave and
discontinuous.
To summarize, Fig.\ref{fig:redgreen}  shows
that  social learning based quickest detection results in fundamentally different decision policies compared to classical quickest time detection (which has a single threshold). Thus making global decisions (stop or continue) based on local decisions (from social learning) is non-trivial.
In \cite{Kri12}, a detailed analysis of the problem is given together with a characterization of this multi-threshold behavior. Also more general
phase-distributed change times are considered in \cite{Kri12}.

\section{Coordination of Decisions in  Sensing -  Non-cooperative Game Approach} \label{sec:game} The
discussion so far has dealt with  Bayesian  social learning models for sensing.
In this section, we present a highly stylized non-Bayesian non-cooperative game theoretic learning approach for adaptive decision making amongst agents.

Social and economic situations often involve interacting decision making with diverging interests. Decision makers may act independently  or form collaborative groups wherein enforceable binding agreements ensure coordination of joint decisions.  For instance, a person may choose the same cellphone carrier as the majority of family and friends to take advantage of the free talk times. Social networks diffuse  information and hence  facilitate coordination of such cooperative/self-interested units. 
This section examines how global coordination of decisions can be obtained when  self interested agents form a social network. 

As mentioned in the Introduction, human-based sensing systems are comprised of agents with partial information and it is the dynamic
interactions between agents that is of interest. This motivates the need for game theoretic learning models for agents interacting in social networks.
Learning dynamics in games typically can be classified into Bayesian learning, adaptive learning and evolutionary dynamics. We have already focussed on 
Bayesian social learning\footnote{The social learning protocol of Sec.\ref{sec:herd} can  be viewed  as a Bayesian game comprising  a countable number of agents,
where each agent plays once
in a specified order to minimize its cost; see \cite{Cha04} for further details on this game-theoretic interpretation.}
 in previous sections of the paper, and some further remarks are made in Sec.\ref{sec:close} on global Bayesian games. 

In this section we focus on adaptive learning
where individual agents deploy simple rule-of-thumb strategies. The aim is to  determine
if such simple individual behaviour can result in sophisticated global behaviour. We are interested in cases where the global behaviour converges to the set
of {\em correlated equilibria}.

\subsection{Correlated Equilibria and Related Work}
The set of correlated equilibria is a more natural construct in decentralized adaptive learning environments than the 
set  of Nash equilibria\footnote{The set of correlated equilibria
is defined in (\ref{ce}) below.  Nash equilibria are a special case of correlated equilibria where
 the joint strategy  is chosen as the product distribution for all players, i.e., all the agents choose their actions  independently.}, since it allows for individual players to coordinate their actions. This coordination can lead to higher performance \cite{Aum87} than if each player chooses actions independently as required by a Nash equilibrium. As described in \cite{HM00}, it is  unreasonable to expect in a learning environment that players act independently (as required by a Nash equilibrium) 
since the common history observed by all players acts as a natural coordination device.\footnote{
Hart and Mas-Colell observe in \cite{HM01b}
that for most simple adaptive procedures,
``...there is a natural coordination device:
the common history, observed by all players.  It is
thus  reasonable to expect that, at the end,
independence among players will not obtain.''}
The set of correlated equilibria is also structurally
simpler than the set of Nash equilibria;  the  set of correlated equilibria is a convex polytope in the set of randomized strategies
whereas  Nash equilibria are isolated points at the
extrema of this set. Indeed, a feasible point in (\ref{ce}) is obtained straightforwardly by a linear programming solver.

\paragraph*{Related works} 
A comprehensive textbook in game theoretic learning is \cite{FL99}.
Algorithms for game-theoretic learning are broadly classified into  best response, fictitious play and regret matching.
In general it is impossible to guarantee convergence to a Nash equilibrium without imposing conditions on the structure of the utility functions in the game.
For supermodular games \cite{Top98},  best response algorithms can be designed to converge either to the smallest or largest Nash equilibrium.
Fictitious play is one of the oldest and best known models of learning in games; we refer the reader to \cite{HS02} for convergence of 
stochastic fictitious play algorithms. In this section we focus on regret-matching algorithms.
Regret-matching as a strategy of play in long-run interactions has been introduced in~\cite{HM00,HM01b}. 
In~\cite{HM00}, it is proved that when all agents share stage actions and follow the proposed regret-based adaptive procedure, the collective behavior converges to the set
of correlated equilibria.
In~\cite{HM01b}, the authors assumed that agents do \emph{not} observe others' actions and proposed a reinforcement learning procedure that converges to the set of 
correlated equilibria.  More recently, \cite{NKY13,NKY13a,KMY08,MKZ08} consider  learning in a dynamic setting where a regret matching type algorithm
tracks a time varying set of correlated equilibria.

 \subsection{Regret-Based Decision Making} 
 
 Consider a non-cooperative  repeated game comprising of $L$ agents.
 Each agent  $l$ has a utility function
$U^l(a^1,\ldots,a^L)$. Here $a^l $ denotes the action  chosen by agent  $l$ and $a^{-l}$ denote the actions chosen by all agents
 excluding agent $l$. The utility function can be quite general. For
example,  \cite{NKY13b} considers the case in which  the $L$ agents are  organized into $M$ non-overlapping social (friendship) groups such that
 agents in a social group share the same  utility function. It could also
 reflect reputation or privacy using the models in \cite{Mui02,GGG09}.

 Suppose each agent $l$ chooses its actions according to the following  adaptive algorithm running over time $k=1,2,\ldots$:
\begin{compactenum}
\item At time $k+1$, choose  action $a_{k+1}^l$ from probability mass function $\psi_{k+1}^l$, where
\begin{multline}
\label{eq:Transition_Matrix_B}
\psi^{l}_{k+1}\left(i\right) = P\left(a_{k+1}^l = i | a_{k}^l\right) \\ = \left\{ \begin{array}{ll}
\frac{ \left|r^{l}_{k}\left(i,j\right)\right|^{+}} {C}, & j\neq a_{k}^l\\
1-\sum_{j\neq i} \frac{ \left|r^{l}_{k}\left(i,j\right)\right|^{+}} { C}, & j=a_{k}^l .\\
\end{array} \right. 
\end{multline}
Here $C$ is a sufficiently large positive constant so that  $\psi^{l}_{k+1}$ is a valid probability mass function.
\item The regret matrix $r_k^l$ that determines the pmf $\psi^{l}_{k+1}$ is updated via the stochastic approximation algorithm
\begin{multline}
\label{eq:SA}
r^{l}_{k}\left(i,j\right) = r^{l}_{k-1}\left(i,j\right)  + \frac{1}{k} \Big(\bigl[ U^l\left(j,{a}_{k}^{-l}\right) \\ - U^l\left(a_{k}^{l},{a}_{k}^{-l}\right) \bigr] I_{\lbrace a_{k}^{l} = i\rbrace} - r^{l}_{k-1}\left(i,j\right)\Big).
\end{multline}
\end{compactenum}
Step 1 corresponds to  each agent choosing its action randomly from a Markov chain with transition probability $\psi^{l}_{k+1}$.
These transition probabilities are computed in 
Step 2 in terms of the regret matrix $r^l_k$ which is the time-averaged regret  agent $l$  experiences for choosing action $i$ instead of action $j$ for each
possible action $j \neq i$ (i.e.,  how much better off it would  be if it had chosen action $j$ instead of~$i$):
\begin{equation}
\label{eq:average_regret}
r^{l}_{n}\left(i,j\right) = \frac{1}{n}\sum_{k = 1}^{n} \left[U^l\left(j,{a}_{k}^{-l}\right)-U^l\left(a_{k}^{l},{a}_{k}^{-l}\right)\right]\cdot I_{\lbrace a_{k}^{l} = i\rbrace}.
\end{equation}

The above algorithm can be  generalized to consider multiple social groups. If agents within each social group share their actions and have a common utility,
then they can fuse their individual regrets into a regret for the social group. As shown in \cite{NKY13b}, this fusion of regrets can be achieved via a linear combination of the individual regrets where
the weights of the linear combination depend on the reputation of the agents that constitute the social group.

\subsection{Coordination in Sensing}
We now address the following question:
\begin{quote} If each agent  chooses its action according to the above  regret-based algorithm, what can one say about the emergent
global behavior?  
\end{quote}
By emergent global behavior, we mean the empirical frequency of actions taken over time by all agents. For each $L$-tuple of actions 
$(a^l,a^{-l})$ define the empirical frequency of actions taken up to time $n$ as
$$ z_n(a^l,a^{-l}) = \frac{1}{n} \sum_{k=1}^n I(a_k = a^l, a_k^{-l} = a^{-l}). $$

The seminal papers \cite{HM00}  and  \cite{Har05} show that  the empirical frequency of actions   $z_n$ converges as $n\rightarrow \infty$ to the set of {\em correlated equilibria} of a non-cooperative 
game.  Correlated equilibria constitute a generalization of Nash equilibria and were introduced by Aumann \cite{Aum87}.
 The set
    of correlated equilibria $\mathcal{C}_e$ is the set of probability distributions on the joint action profile $(a^l,a^{-l})$  that satisfy
    \begin{multline}
    \label{ce}
\mathcal{C}_e = \biggl\{\mu:
    \sum_{a^{-l}} \mu^l(j,a^{-l})[U^l((i,a^{-l}))-U^l((j,a^{-l}))] \\ \leq 0, 
    \quad \forall l,j,i\biggr\}.
    \end{multline}
Here  $\mu^l(j,a^{-l}) = P^l(a^l = j, a^{-l})$ denotes the randomized strategy (joint probability) of player $l$ choosing action $j$ and the rest of the players choosing action $a^{-l}$.
The correlated equilibrium condition (\ref{ce}) states
that instead of taking action $j$ (which is prescribed by the equilibrium strategy $\mu^l(j,a^{-l})$), if player $l$ cheats and takes action $i$, it is worse off. So there is no unilateral incentive for any player  to cheat.

To summarize, 
the above  algorithm ensures that all agents eventually achieve {\em coordination (consensus) in decision making} -- the randomized strategies
 of all agents  converge to a common convex polytope $\mathcal{C}_e$. 
 Step 2 of the algorithm requires that each agent knows its own utility and the actions of other agents -- but agents do not need to know the utility
 functions of other agents.  In \cite{HM01b} a `blind' version of this
 regret-based algorithm is presented in which agents do not need to know the actions of other agents.
These  algorithms
 can  be viewed as  examples in which simple heuristic behavior by individual agents (choosing actions according to the measured regret) resulting in sophisticated  global outcomes \cite{Har05}, namely
 convergence to $\mathcal{C}_e$ thereby  coordinating decisions.
 
We refer to \cite{KMY08,MKZ08,NKY13} for generalizations of the above algorithm
to the tracking case where the step size for the regret matrix
 update is a constant.
 Such algorithms can track the correlated equilibria of games with time-varying parameters. Moreover  \cite{NKY13a} gives
 sufficient conditions for algorithm to converge to the set of correlated equilibria when the  regrets from agents to other agents diffuse over a social network.


\section{Closing Remarks} \label{sec:close}
In this paper we have used social learning as a model for interactive sensing with social sensors.
We summarize here some  extensions of the social learning framework that are relevant to interactive sensing.
\subsubsection{Wisdom of Crowds} 

Surowiecki's  book \cite{Sur05}  is an excellent popular piece that 
explains the wisdom-of-crowds hypothesis. The wisdom-of-crowds hypothesis predicts that the independent judgments of a crowd of individuals (as measured by any form of central tendency) will be relatively accurate, even when most of the individuals in the crowd are ignorant and error prone. The book also 
studies situations (such as rational bubbles) in which  crowds are not wiser than individuals.  Collect enough people on a street corner staring at the sky, and everyone who walks past will look up. Such herding behavior is typical in social learning.

\subsubsection{In which order should agents act?}
In the social learning protocol, we assumed that the agents act sequentially in a pre-defined order.
However, in many social networking applications, it is important to optimize the order in which agents  act. For example, 
consider an online review site where individual reviewers with different reputations  make their reviews publicly available. 
If a reviewer with high reputation publishes its review first, this review will unduly affect the decision of a reviewer with lower reputation.
In other words, if the most senior agent ``speaks" first it would unduly affect the decisions of more junior  agents. This could lead to an increase in bias of the underlying state estimate.\footnote{To quote a recent paper from Haas School of Business, U.C. Berkeley \cite{AK09}: ``In  94\% of cases, groups (of people) used the first answer provided as their final answer... Groups tended to commit to the first answer provided by
any group member.''  People with dominant personalities tend to speak first and most forcefully ``even when they actually lack competence''. } 
On the other hand, if the most junior agent is polled first, then since its variance is large, several agents would need to be polled in order
to reduce the variance. We refer the reader to \cite{OS01} for an interesting description of who should speak first in a public debate.\footnote{As described
in \cite{OS01}, seniority is considered in the rules of debate and voting in the U.S.\ Supreme Court. ``In the past, a vote was taken after the newest
justice to the Court spoke, with the justices voting in order of ascending seniority largely, it was said, to avoid the pressure from long-term members
of the Court on their junior colleagues."}
It turns out that for two agents, the seniority rule is always optimal for any prior -- that is, the  senior agent speaks first followed
by the junior agent; see \cite{OS01} for the proof. However, for more than two agents, the optimal order
depends on the prior, and the observations in general.

\subsubsection{Global Games for Coordinating Sensing}  In the classical Bayesian  social learning model of
Sec.\ref{sec:classicalsocial}, agents act sequentially in time. The global games model that has been studied
in economics during the last two decades, considers multiple agents that act simultaneously by predicting the behavior
of other agents. 
The theory of global games was first introduced in \cite{CD93} as a tool for refining equilibria in economic game theory;
see \cite{MS00} for an excellent exposition.
Global games
are an ideal method for decentralized coordination amongst agents;   they 
have  been used to model speculative currency attacks and regime change in social systems,  see
\cite{MS00,KLM07,AHP07}.

 The most widely studied form of a global game is a one-shot Bayesian game which proceeds as follows: Consider a  continuum of agents  in which each agent $i$ obtains noisy measurements $y^i$
of  an underlying state of nature $x$. Assume all agents have the same
observation likelihood  density $p(y|x)$ but the individual measurements obtained by agents
are statistically independent of those obtained by other agents. Based on its observation $y^i$, each agent 
takes an action $a^i \in \{1, 2\}$ to optimize its expected utility $\E\{U(a^i, \alpha) | y^i\})$ where
$\alpha \in [0,1]$ denotes the fraction of all agents that take action 2.  Typically, the utility $U(1,\alpha)$ is set to zero.

For example,  suppose $x$  (state of nature) denotes the quality of a social group and  $y^i$ denotes the measurement of this quality by agent $i$.  The action $a^i = 1$  means that agent $i$ decides not to  join the social group, while  $a^i = 2$ means that agent $i$ joins the group.
The utility function $U(a^i=2,\alpha)$ for joining the social group depends on $\alpha$, where $\alpha$ is the fraction of people that decide to join the  group. 
In  \cite{KLM07}, the utility function is chosen as follows:
If $\alpha \approx 1$, i.e.,  too many people join the group, then  the utility to each agent is small since the group is too congested and agents do not receive sufficient individual service.
On the other hand, if $\alpha \approx 0$, i.e.,  too few people join the group, then the utility  is also small since there is not enough social interaction.

Since each agent is  rational, it  uses its  observation $y^i$ to predict $\alpha$, i.e., the fraction of  other agents 
that choose action 2. The main question is: What is the optimal strategy for each agent $i$ to maximize its expected utility?

It has been shown that for 
a variety of 
  measurement noise models (observation likelihoods $p(y|x)$) and  utility functions $U$, the 
 symmetric Bayesian 
Nash equilibrium of the global game is unique and 
has a threshold structure in the observation. This means that given its observation $y^i$, it is optimal for each agent $i$ to choose
 its actions as follows:
 \begin{equation} \label{eq:thresbne}
  a^i = \begin{cases} 1 &  y^i < y^* \\ 2  & y^i \geq y^* \end{cases}
 \end{equation}
where the threshold $y^*$  depends on the prior, noise distribution and utility function.

In the above example of joining a social group, the above result means that if agent $i$ receives a   measurement $y^i$ of the quality of the group, and $y^i$  exceeds a threshold $y^*$, then it should join.
This is yet another example of simple local behavior (act according to a threshold strategy) resulting in global sophisticated behavior (Bayesian Nash equilibrium).
As can be seen, global games provide a decentralized way of achieving coordination amongst  social sensors.
In \cite{AHP07}, the above one-shot Bayesian game is generalized to a dynamic (multi-stage)  game operating over a possibly infinite horizon. Such games facilitate modelling the dynamics of how people join, interact and leave
social groups.

  The papers \cite{Kri08,Kri09} use global games to model networks of sensors and cognitive radios.
In \cite{KLM07} it has been shown that the above threshold structure (\ref{eq:thresbne}) for the Bayesian Nash equilibrium, breaks
down if the utility function $U(2,\alpha) $  decreases too rapidly due to congestion. The equilibrium structure becomes much more complex
and  can be described by the following  quotation \cite{KLM07}:
\begin{quote} {\em Nobody goes there anymore. It's too crowded} -- Yogi Berra \end{quote}



\section{Summary} \label{sec:extension}
This paper has considered  social learning models for interaction among  sensors where agents use their private observations along with actions of other agents to estimate an underlying state of nature.
We have considered extensions of the basic social learning paradigm to online reputation systems in which agents communicate over a social network.
 Despite the apparent simplicity in these information flows, the systems exhibit unusual behavior such as herding and data incest.
 Also an example of social-learning for change detection was considered.
Finally, we  have discussed  a non-Bayesian formulation, where agents seek to achieve coordination in decision making by optimizing their own utility functions - this was formulated as a game theoretic
learning model. 

 The motivation for this paper stems from understanding how individuals interact in a social network and how simple local behavior can result in complex global behavior.
The underlying  tools
used in this paper are widely used by  the electrical engineering research community in the areas of signal processing, control, information theory and network communications.

\bibliographystyle{IEEEtran}\bibliography{$HOME/styles/bib/vkm}

\begin{IEEEbiographynophoto}{Vikram Krishnamurthy}
[F]  (vikramk@ece.ubc.ca)   is  a professor and Canada Research Chair at the
Department of Electrical Engineering, University of British Columbia,
Vancouver, Canada. Dr Krishnamurthy's  current research interests include statistical signal processing, computational game theory and
stochastic control in social networks. He served as distinguished lecturer for the IEEE Signal Processing Society and
Editor in Chief of IEEE Journal Selected Topics in Signal Processing. He received an honorary doctorate from KTH (Royal Institute of Technology), Sweden
in 2013.
\end{IEEEbiographynophoto}
\begin{IEEEbiographynophoto}{H. Vincent Poor}
[F] (poor@princeton.edu) is Dean of Engineering
and Applied Science at Princeton University, where he is also the 
Michael Henry Strater University Professor. His interests include 
statistical signal processing and information theory, with applications 
in several fields. He is a member of the National Academy of Engineering,
the National Academy of Sciences, and the Royal Academy of Engineering (UK). 
Recent recognition includes the 2010 IET Fleming Medal, the 2011 IEEE
Sumner Award, the 2011 Society Award of IEEE SPS, and honorary 
doctorates from Aalborg University, the Hong Kong University of 
Science and Technology, and the University of Edinburgh. 
\end{IEEEbiographynophoto}

\vfill
\end{document}